\documentclass[12pt,preprint]{aastex}
\usepackage{amssymb,amsmath,float}

\begin{document}

\title{The Sunyaev-Zel'dovich effect in a sample of
31 clusters - a comparison between the 
X-ray predicted and WMAP observed CMB temperature decrement}

\author{Richard Lieu$\,^{1}$,
Jonathan P.D. Mittaz$\,^{1}$, and
Shuang-Nan Zhang$\,^{2,1,3,4}$
}

\affil{\(^{\scriptstyle 1} \){Department of Physics, University of Alabama,
Huntsville, AL 35899.}\\
\(^{\scriptstyle 2} \){Physics Department and Center for
    Astrophysics, Tsinghua University, Beijing, 100084, China.}\\
\(^{\scriptstyle 3} \){Key Laboratory of Particle Astrophysics,
Institute of High Energy Physics, Chinese Academy of Sciences,
P.O. Box 918-3, Beijing 100039, China.}\\
\(^{\scriptstyle 4} \){NASA
Marshall Space Flight Center, NSSTC, SD-50, 320 Sparkman Drive,
Huntsville, AL 35805}
}

\begin{abstract}
The WMAP Q, V, and W band radial profiles of temperature deviation of the
cosmic microwave background (CMB) were constructed for a sample of 31 randomly
selected nearby clusters of galaxies in directions of Galactic latitude $|b| >$
30$^o$.  The profiles were compared in detail with the expected CMB
Sunyaev-Zel'dovich effect (SZE) caused by these clusters, with the hot gas
properties of each cluster inferred observationally by applying
gas temperatures as measured by ASCA to
isothermal $\beta$ models 
of the ROSAT X-ray surface brightness profiles, and with
the WMAP point spread function fully taken into consideration.
A reliable overall assessment can be
made using the combined (co-added) datasets of all 31 clusters, because (a) any
remaining systematic uncertainties are low, and (b) the data are extremely
clean (i.e. free from foreground contaminants).  Both (a) and (b) are facts
which we established by examining hundreds of random fields.  
After co-adding the 31 cluster field, it
appears that
WMAP detected the SZE in all three bands.   Quantitatively, however,
the observed SZE
only accounts for about
$1/4$ of the expected decrement.  The discrepancy represents too much 
unexplained extra
flux: in the W band, the detected SZE corresponds on
average to 5.6 times less X-ray gas mass within a 10 arcmin radius than
the mass value given by the ROSAT $\beta$ model.  We examined critically
how the X-ray prediction of the SZE may depend on our uncertainties
in the density and temperature of the hot intracluster plasma,
and emission by cluster radio sources.  Although our comparison between
the detected and expected SZE levels 
is subject to a margin of error,
the fact remains
that the average observed SZE depth and profile are consistent with those
of the primary CMB anisotropy, i.e. in principle the average WMAP temperature
decrement among the 31 rich clusters is too shallow to 
accomodate any extra effect like the SZE.
A unique aspect of this SZE
investigation is that because all the data being analyzed are in the
public domain, our work is readily open to the scrutiny of others.
\end{abstract}

\noindent
{\bf 1. Introduction}

\noindent
One vital test of the present cosmological paradigm is
the search for
scattering of the CMB
by foreground structures such as clusters of galaxies.  Such
observations can provide important information both about clusters of galaxies
as well as basic cosmological parameters like $H_0$.
For the CMB, scattering arises from the Compton interaction with free
electrons in the hot (X-ray temperature) plasma of clusters of galaxies, which
removes Rayleigh-Jeans black body flux in the direction to a cluster and leads
to an apparent decrease in the CMB temperature - a phenomenon known as the
Sunyaev-Zel'dovich effect (SZE).  By now, the degree of SZE is highly
predictable for many clusters of galaxies, because their hot intracluster
medium (ICM) properties are well measured by X-ray satellite missions.

In this work we propose to perform the just such a detailed
comparison.  Of course, among the earlier papers that involved satellite data
(see section 6 on interferometric techniques), the X-ray morphology and spectra
of many clusters were already investigated in depth, while marginal detections
of the SZE by WMAP were reported for individual and entire ensemble of clusters
(Bennett et al 2003a; Myers et al 2004; Hernandez-Monteagudo, C.,
Genova-Santos, R., \& Atrio-Barandela, F. 2004; Hernandez-Monteagudo, C., \&
Rubino-Martin, J. A. 2004; Afshordi, Lin, \& Sanderson 2005).
Nonetheless, our work in this paper represents the first time in which
SZE {\it in radial profiles of fluxes with errors},
rather than false color significance maps,
are shown in three separate passbands for
individual clusters within a large (31 member) sample, leading to
an in-depth evaluation of whether the WMAP SZE profiles are consistent
with the hot ICM profiles and temperatures as measured by X-ray missions.

\vspace{2mm}

\noindent
{2. \bf The cluster sample and predicted SZE profiles from X-ray data}

The sample employed for our purpose is the Bonamente et al (2002) catalog of 38
nearby clusters of galaxies (hereafter simply referred to as the Bonamente
sample) located in directions of high Galactic latitude ($|b| > 30^{o}$) and
low column density of Galactic neutral hydrogen.  The rationale has to do with
our original intention of using the SZE as a sensitive probe of any extra
baryonic matter that may exist in the outskirts of clusters in the form of a
warm gas.  Since this aim led eventually to a
different and much more surprising finding we
will no longer discuss it, except to mention that there is no reason why the
members of the Bonamente list should constitute a biased sample in any way, as
far as their utility as a probe of the CMB distance scale is concerned.  To the
contrary, their high $|b|$ location means that the WMAP data for these clusters
have minimal Galactic foreground contamination problems.

In order to assess whether the temperature feature recorded by WMAP at the
position of these clusters is consistent with SZE from the X-ray emitting hot
ICM, it is necessary to first determine the radial distribution of the hot ICM
density of each cluster using the data of X-ray observatories.  In particular,
we employed the ROSAT database, because the ROSAT XRT has a sufficiently large
field-of-view to enable background determination even for the very extended
nearby clusters in the sample, such as Coma.  For clusters without a `cooling
flow', the standard isothermal $\beta$-model is fitted to the X-ray surface
brightness $I_X (\theta)$ in the entire region from the X-ray centroid $\theta=
0$ to the background.  Specifically three parameters are fixed by modeling the
surface brightness profile:

\begin{equation}
I_X (\theta) \propto n_0^2 \left[ 1 + \left(\frac{\theta}{\theta_c}\right)^2 
\right]^{-3\beta +\frac{1}{2}},
\end{equation}
where $\theta_c$ is the core radius, $n_0$ is the central electron number
density, and $\beta$ is the decay index.  To predict the SZE another cluster
parameters, the temperature $T$ of the clusters, is also needed, and
these were taken from Bonamente et al (2002).  Where possible the temperatures
were measured by ROSAT, though ASCA data were used when a cluster's ICM is too
hot for ROSAT's passband.  With the advent of Newton and Chandra, the reported
temperatures corroborate ASCA but are generally slightly higher than the ROSAT
values.  Since the SZE depth enhances with increasing $T$, the Bonamente
temperatures are in fact too conservative for estimating this depth.

For `cooling flow' clusters, the central region where cooling occurs is
excluded from the modeling.  The repercussions of such a procedure are
discussed at length in section 5, where high resolution XMM-Newton data
will be used to demonstrate that any errors in the SZE associated with
this analysis method are too
small to affect our final conclusions.  All
the ROSAT radial profiles of the member clusters of the Bonamente
sample are shown in Figure \ref{rosat}.  The best-fit $\beta$-model parameters,
together with other essential details of each cluster, are to be found in Table
1, where a spot check against Briel \& Henry (1996) reveals reasonable
agreement with their values of $n_0 =$ 3.15 $\pm$ 0.29 $\times$ 10$^{-3}$
cm$^{-3}$, $\theta_c =$ 5.15 $\pm$ 0.46 arcmin, and $\beta =$ 0.93 $\pm$ 0.04
for A1795.

The predicted SZE decrement
as a function of angle relative to the cluster center
direction is then given by
\begin{equation}
\frac{\Delta T_{{\rm SZ}} 
(\theta)}{T_{{\rm CMB}}} = \frac{kT}{m_e c^2} \sigma_{{\rm Th}}
\int dl~n_e \left[\frac{x(e^x +1)}{e^x -1} - 4 \right],
\end{equation}
where $x = h\nu/kT_{{\rm CMB}}$ with $\nu$ being the mean frequency of the WMAP
observing filter, $\sigma_{{\rm Th}}$ is the Thomson cross section, $l$ is the
pathlength through the cluster along our off-axis sightline, and the electron
density is
\begin{equation}
n_e(r) = n_0 \left[1 + \left(\frac{r}{r_c}\right)^2 \right]^
{-\frac{3\beta}{2}}, 
\end{equation}
with $r = r(l, \theta)$ being the radius.  The integration of Eq. (2) was
performed analytically by previous authors, resulting in an expression for
$\Delta T_{{\rm SZ}} (\theta)/T_{{\rm CMB}}$ 
which depends on $n_0$, $\beta$, and the core
radius $r_c = L\theta_c$ where $L = cz/H_0$ (with $H_0 =$
71 km s$^{-1}$ Mpc$^{-1}$) is the distance to a 
nearby cluster, see
e.g. Refregier, Spergel, \& Herbig (2000).  We will not repeat these earlier
calculations, except to say that upon application of our Coma cluster
$\beta$-model parameters to the analytical formula we obtained 
$\Delta T_{{\rm SZ}}
(\theta=0)/T_{{\rm CMB}} \approx -$590 $\pm$ 170 $\mu$K at the WMAP frequency
of $\nu =$ 41 GHz.  This compares well with the estimate by other groups, such
as the value of 
$\Delta T_{{\rm SZ}} (\theta=0)/T_{{\rm CMB}} \approx -$507 $\pm$ 92.7
$\mu$K at 32 GHz as obtained by Herbig, Lawrence, \& Readhead (1995).

\vspace{2mm}

\noindent
{\bf 3.  The cluster WMAP temperature profiles; effect of the
point spread function}

To examine whether the WMAP mission detected temperature variation in the
fields of the Bonamente sample we extracted `thumbnail' images (small maps)
from the WMAP first year database centered at the X-ray centroids of the 38
clusters, i.e. the coordinates as listed in Table 1.  Clusters with bright
radio sources along their sightlines (Virgo, A21, A1045) are excluded from
further analysis.  In addition, Fornax, A1314, and A1361 were excluded due to
difficulties in estimating the uncertainty in the X-ray brightness
distribution.  Lastly, Hercules was not considered because the exact location
of the cluster emission could not be identified unambiguously.  For the
remaining 31 sample members radial profiles of the mean temperature deviation
over concentric annuli are computed after removing the dipole and quadrupole
components, and plotted in Figure \ref{wmap}, where the error bars for each
radius interval reflect an antenna noise which scales as the
inverse-square-root of the number of observations of the sky area.  We restrict
attention to the cosmological bands of Q, V and W, since the K and Ka bands are
dominated by Galactic foreground emissions and absorptions (Bennett et al
2003a,b).

In order to compare any spatial features seen at a cluster
position with the expected  SZE behavior, it is necessary to take
into account the WMAP point spread function (PSF).  For each of the three
filters we computed the average brightness profile from 15 point sources
and fitted it with a Gaussian function as shown in Figure \ref{wmap-psf}.   
It was possible to perform the averaging of these profiles because they
all compare well with each other statistically, even though the sources
are located at very different parts of the sky, with the highest
being at Galactic latitude
$|b| \approx$ 74$^o$.  This indicates that the performance of
WMAP in spatially resolving CMB temperature variations is consistent
across the sky.
The SZE profile of each cluster as inferred from Eq. (2) and 
Table 1 is
convolved with the PSF of the appropriate filter.  A test of
the correctness of our procedure was done by adopting the
$\beta$-model in  Figure 1(c) of Myers et al (2004).  After
convolving this model with the W-band PSF, our results are plotted
in Figure \ref{shanks}.  It bears 
close resemblance to the corresponding profile
in the Myers paper.  

With the assurance of the aforementioned cross-checks the
expected SZE decrement for each cluster is seen overplotted in
Figure \ref{wmap}, where the unperturbed `continuum' is aligned
with the average temperature deviation in
the outermost (2$^\circ$ - 3$^\circ$) annulus.
The {\it co-added} model and data profiles for each filter is
shown in Figure \ref{sze-prediction}.   In all plots 
the error bars for each radius interval reflect an antenna
noise which scales as the inverse-square-root of the number of observations
of the sky area.  An analysis of the (radial) bin-to-bin variation of
the profile of many co-added {\it random} fields
reveal that the degree
of relative fluctuation between
neighboring radial bins is in accordance with the size of the error
bars as shown in Figure \ref{sze-prediction} for each bin.
This enables us
to evaluate the formal statistical significance of the overall detection
of a broad temperature decrement feature in the composite radial profiles
as 9$\sigma$,
4.2$\sigma$, and 2.3$\sigma$ for the Q, V, and W band respectively (see
Figure \ref{sze-prediction} for more details).
Note that the true significances are likely to be
less than those given by the three numbers, because
systematics effects in the WMAP data were not included when we
calculated them.
Such effects will
be discussed at length in the next section.

\vspace{2mm}

\noindent
{\bf 4.  Systematic CMB temperature variation from WMAP random fields}

From the graphs in Figure \ref{wmap}, a 
broad CMB temperature decrement positionally coincident with the
cluster and of commensurate spatial extent as the cluster size is
apparent for some clusters, such as
Coma, A1795, A1413, A2199, A2219, and A2255.  On the other hand, there
are clearly counter examples like A85, A1367, A1689, A2029, A3301,
A3558, and A3571.  There are also clusters that fall within a
`grey' zone where no judgment can be made at all.  Apart from
some clusters having small expected SZE amidst poor signal-to-noise, another
key problem that weakens any verdict from an individual cluster
is the ambiguity in the CMB temperature `continuum' appropriate
to the observation.  In fact, we are dealing with a non-trivial systematic
uncertainty which can only be understood by examining many randomly
chosen fields across the WMAP sky with $|b| \geq$ 30$^o$.  For this reason
we constructed radial profiles for 100 of such fields.

A plot of the r.m.s. variation (from one random field to another)
of the temperature difference at each radial interval is
shown in Figure \ref{random-stdev}, where it is evident that systematic effects
at the level
of 0.1 mK are commonplace among the smaller radius annuli.  To be 
even more specific, 
we display in Figure \ref{random-single} 
the radial profiles of four random fields.
It can be seen that degree-scale modulations are frequently present,
with an amplitude of $\approx$ 0.1 mK, often found at the $\theta =$ 0
(i.e. on-axis) position.
The phenomenon pertains to the most
prominent fluctuation in the WMAP data - the primary acoustic peak -
which has an amplitude of $\Delta T_{{\rm CMB}} \approx$ 0.1
(or $\Delta T_{{\rm CMB}} /T_{{\rm CMB}} \approx$ a few $\times$
10$^{-5}$, see Bennett et al 2003b).  Thus, when a single cluster
field exhibits some broad central `bump' it is not
necessarily the signature of an emission source; in fact we shall demonstrate
that {\it for our cluster sample as a whole}
the discrepancy between predicted and detected SZE profiles cannot
be due to line-of-sight emissions.  Likewise, the existence of a
central trough may also be caused by 
primordial acoustic oscillations rather
than the SZE.  The only way of knowing whether there is consistency
between the WMAP data and cluster SZE's is to examine the {\it co-added}
profiles of all the clusters, when the systematic effects which prevent
one from determining the `continuum' level can largely be suppressed.

In order to be certain that the temperature does stabilize among co-added
fields we generated average radial profiles of 33 random fields at a time
(Figure \ref{random}).  If the primary acoustic peaks and troughs are stacked
with arbitrary central alignment in this way, one would expect the fluctuation
amplitude to be reduced from the $\sim$ 0.1 mK value by a factor of $\approx
\sqrt{30}$, to about 0.02 mK.  This is completely borne out by our averaged
random profiles, which also indicate that towards the larger 
(2$^\circ$ -- 3$^\circ$)
annuli the CMB temperature deviations from the global mean value are very
uniform, hovering close to zero.  By the time 100 randomly generated profiles
are averaged, the data points are smooth throughout all radii (Figure
\ref{random}).  With the help of these plots (of the merged 100 fields),
wherein the systematic variations are no longer a problem, we could verify if
the error bar on each radial bin (antenna random noise) is independent of other
bins.  The answer is yes, because the r.m.s.  scatter among the data points is
found to be consistent with the size of the error bars.

Returning now to the averaging of 33 random fields, Figure \ref{random},
the residual systematic
excursions in the data are important towards understanding
the radial profile of the co-added cluster fields, Figure 
\ref{sze-prediction}.  This
is because both Figures involve the stacking of a similar number of
WMAP fields.  Since the magnitude of any remaining systematic effect is,
as can be seen in Figure \ref{random},
far less than the discrepancy between observed
and expected SZE in our cluster sample,
Figure
\ref{sze-prediction}, one must conclude that
{\it the apparent incomplete SZE
is not due to intrinsic sky variations in the
CMB temperature recorded by WMAP,} i.e. it is a genuine anomaly that
deserves an explanation.

\vspace{2mm}
 
\noindent
{\bf 5.  Interpretation}

How can one reconcile a cosmological CMB origin
with Figure  \ref{sze-prediction}?  It is perhaps more reasonable to
first examine whether,
despite many generations of
X-ray observatories measuring the X-ray properties of clusters,
we are still misled by uncertainties on the
hot ICM parameters?  Then there
is the question of radio source contamination also.

{\it Could a steepening in the slope of the hot ICM density profile
beyond the radius where the $\beta$-model 
is well constrained by ROSAT data lead
to an overestimation of the SZE?}  We pause to consider an extreme scenario
under which the WMAP's spatial resolution is so poor that the entire SZE
of even a nearby cluster simply appears as a `point sink' at zero
radius.  If this is really the case, the WMAP radial profile of the SZE would 
have the shape of the PSF, with the flux within the entire profile being
equal to the cluster SZE integrated over all sightlines cutting
through the cluster at various `impact parameters'.  For a $\beta$-model,
however, this total flux is dominated by the SZE at the outer radii
where the model is no longer well constrained by the X-ray data.  The
reason is that at any off-axis angle $\theta$ the line-of-sight SZE has
the form:
\begin{equation}
\Delta T_{{\rm SZ}} (\theta) = \Delta T_{{\rm SZ}} (0) \left[ 1 +
\left(\frac{\theta}{\theta_c}\right)^2 \right]^{-\frac{3\beta}{2} 
+ \frac{1}{2}}.
\end{equation}
At least for our sample of clusters ROSAT data fail to guide the model
only at radii $\theta \gg \theta_c$, and typically at $\theta > $ 10 arcmin
in real units.  

The total X-ray predicted SZE, integrated over all values of 
$\theta$, is
dominated by the values of $\Delta T_{{\rm SZ}} (\theta)$ in the
range $\theta \gg \theta_c$, unfortunately.  This is because 
\begin{equation}
\Delta T_{{\rm SZ}}^{{\rm total}} = \int_0^\infty \Delta T_{{\rm SZ}} (\theta)~
2\pi\theta d\theta,
\end{equation}
and for $\theta \gg \theta_c$ the integral $\sim \theta^{-3(\beta - 1)}$,
which diverges at the upper integration limit because the inequality
$\beta <$ 1 applies to our clusters (see Table 1).  In fact, from Eq. (3)
it can also be seen that the total cluster mass, proportional to
$\int n_e(r) r^2 dr$, scales with the upper cutoff radius in exactly
the same way.  Thus the outcome of our analysis is that {\it if} WMAP's
spatial resolution is very poor the X-ray model can overpredict the
SZE by an arbitrarily large amount.
This point was already raised in the
recent papers of Benson et al (2004) and Schmidt, Allen, and Fabian (2005).
Specifically, while Schmidt, Allen, and Fabian (2005) reported a steepening 
in the slope of the $\beta$-model (i.e. less hot ICM) towards a cluster's
outskirts by comparing Chandra and ROSAT data,
Benson et al (2004) found that provided the SZE is evaluated over a radius
commensurate with the resolution (or beam size) of the instrument any
difference among the predictions derived from the various sets of model
parameters actually remains small.   

In the context of this last point of
Benson et al (2004), we shall demonstrate that most of the
results we obtained in the previous sections remain robust: the WMAP
resolution is not as pessimistic as that depicted in the extreme scenario
we just considered, because the detected SZE profiles are much wider than
the instrument PSF.  In fact, much of the SZE within the
central 0.5$^{\circ}$ radius of
`matching' with the WMAP beam is not
due to the PSF leaking signals from large radii inwards.
Instead the opposite is true: there is a loss of inner signals outwards
which only widens the actual gap between SZE prediction and observation.
To test these statements, we chose a pair of $\beta$-model parameters
typical to our cluster sample, viz. $\beta = 2/3$ and $\theta_c =$ 2 arcmin,
and truncate the full line-of-sight integrated SZE profile,
Eq. (4), abruptly at
$\theta =$ 10 arcmin, resulting in the following functional dependence: 
\begin{equation}
\Delta T_{{\rm SZ}}^{{\rm cutoff}} (\theta) = \Delta T_{{\rm SZ}} (0) [1 +
0.25 (\theta/{\rm arcmin})^2]^{-\frac{1}{2}}~{\rm for}~\theta \leq
10~{\rm arcmin};~{\rm and}~= 0~{\rm otherwise.}
\end{equation}
The reason for setting the cutoff at 10 arcmin is that for
most of the clusters in our sample, Figure \ref{rosat} indicates the ROSAT
data can constrain the surface brightness profile out to at least
such a radius.   If, after
convolution of the truncated profile with the PSF there is a significant
reduction of the inner SZE, this would imply a severe flux
overprediction problem at small radii by the WMAP
PSF, which spreads signals inwards from regions
beyond 10 arcmin where the $\beta$-model is no longer so reliable.  Note
that in our test we did not truncate the $\beta$-model self-consistently
by reducing the line-of-sight integration to reflect the projection effect
of the cutoff at $\theta \leq$ 10 arcmin.  This is a separate question
to be investigated below.

The outcome of the test is shown in Figure \ref{convergence}, 
where it can be seen
that within $\theta \leq$ 0.5 arcmin the reduction in SZE
by our truncation procedure is $\approx$ 12 \% for the Q band, and
much less for the V and W bands.  This  means any incorrectness in the
outer (questionable) parts of the $\beta$-model, when coupled with
the WMAP PSF spreading effect, cannot lead to an overprediction of the
SZE within the central 0.5 arcmin radius by a factor of four to six.
Yet such factors are necessary to explain the average discrepancy between
X-ray expectation and WMAP observed SZE levels for the three cosmological
filter passbands, see Figure \ref{sze-prediction}.   We must also point out that in terms
of PSF effects the real problem is the opposite, viz.
the loss of inner signals to the
radii beyond, as borne out by the fact that when we convolved 
$\Delta T_{{\rm SZ}}^{{\rm cutoff}} (\theta)$ with the PSF and
compared the total SZE over $\theta \leq$ 10 arcmin with the same
quantity obtained before convolution, the ratio of the latter to the
former ranges from 2.48 (Q band) through 5.31 (V band) to 5.61 (W band).  Since
the average X-ray predicted SZE
profile is more centrally peaked than the 
WMAP observed profile, Figure \ref{sze-prediction},
this means the discrepancy between ROSAT and WMAP should have been even more
pronounced if the original {\it intrinsic} profiles before PSF convolution
were compared.

In order to assess the line-of-sight projection effect of the outer parts of 
the $\beta$-model on the inner SZE prediction, we  may e.g.
divide the central decrement $\Delta T_{{\rm SZ}} (0)$ into
two parts, respectively contributions from the hot ICM within and beyond
the truncation limit of
$\theta = \theta_T =$ 10 arcmin,
i.e. we write
\begin{equation}
\Delta T_{{\rm SZ}} (0) \propto \int_0^{\theta_T} 
\left[ 1 + \left(\frac{\theta}{\theta_c}
\right)^2 \right]^{-\frac{3\beta}{2}} d\theta + \int_{\theta_T}^{\infty}
\left[ 1 + \left(\frac{\theta}{\theta_c} \right)^2 \right]^{-\frac{3\beta}{2}} 
d\theta.
\end{equation}
For our typical sample parameters
of $\beta =$ 2/3 and $\theta_c =$ 2 arcmin
the ratio of the 2nd term on the right side
to the first term is 14.5 \%.  
It is clearly not of a magnitude large enough to explain the
discrepancy of concern, which is at the 400 -- 600 \% level
(Figure \ref{sze-prediction}).  
The same conclusion may be drawn about the other
slightly off-axis sightlines of the innermost 0.5$^\circ$ radius.

The impact of uncertainties in the X-ray $\beta$-model on the integrity
of the present work
may quantitatively be summarized into two points.  Firstly, we can
calculate the significance of the average discrepancy between the
predicted and
observed SZE as depicted in Figure \ref{sze-prediction}, taking into
account: (a) the random error bars shown in
Figure \ref{sze-prediction} (i.e. antenna noise for each radial bin, see
the end of section 3); (b) the systematic error per bin which is also shown
in Figure \ref{sze-prediction} and described in section 4; (c) the
uncertainty in the $\beta$-model parameters manifested as errors in the
model prediction for each bin, shown in Figure \ref{sze-prediction} as
the vertical interval between the dashed and solid lines; and (d) 
systematic error in the $\beta$-model prediction for the inner radial bins
due to unreliability of the model at the cluster outskirts coupled with the
WMAP PSF smearing effect, as described in
this section and quantified in Figure \ref{convergence}.
When all four error components are added in quadrature,
we find that for the central 1 arcmin radius the discrepancy between
prediction and observation has the significance level of 2.73 $\sigma$, 
4.65 $\sigma$, and 4.70 $\sigma$ respectively for the Q, V, and W band.
This level becomes even higher if the next few bins are also included
with our calculation.

Onto our second summary point.  As
was argued earlier, since the total SZE summed over all sightlines out
to some limiting radius $\theta \gg \theta_c$ scales with $\theta$ as
the total hot ICM mass (out to the same radius) does, we used 
Figure \ref{sze-prediction}
to compute an ensemble average over our 31 member cluster sample
the ratio of the X-ray predicted to WMAP observed total SZE out to
$\theta =$ 10 arcmin (i.e. $\theta \gg \theta_c \approx$ 2 arcmin).
This ratio, which ranges from 1.49 (Q band) through 2.21 (V band) to
2.82 (W band), indicates by how much must the X-ray $\beta$-model
estimate of the hot ICM mass within the radius $\theta =$ 10 arcmin
be reduced in order to secure a match between ROSAT and WMAP.
For the W band, which is the best (cleanest) cosmological filter
of WMAP, the answer is close to 300 \%.  It is somewhat surprising that
the $\beta$-model can lead to such large errors in the X-ray gas mass: errors
which effectively deemed many previous X-ray observations 
of the hot ICM as meaningless.
Note also that, as
already explained, because the
X-ray predicted SZE profile is more centrally peaked than the WMAP
profile, problems with the leakage of flux towards the $>$ 10 arcmin radii by
the WMAP PSF render the correct percentage reduction of the hot ICM
mass even higher than 300 \%.

{\it Could a `cooling flow' which exists in some clusters invalidate any
estimate of central SZE based upon isothermal $\beta$-model fits to the
ROSAT data?}   While remaining in the spirit of the foregoing discussion, viz.
on the question of how
the SZE predicted profile for the central 0.5$^{\circ}$ radius of
`matching' with the WMAP beam  may be
affected by the on-axis flux decrement, we turn to the
possible role played by the cooling of cluster cores.  By using the more
accurate measurements of Chandra, Schmidt, Allen, and Fabian (2005) found
that a straightforward ROSAT $\beta$-model inference of 
$\Delta T_{{\rm SZ}} (0)$ could lead
to an overestimate of the quantity by less than a factor of two.  Apart from
re-emphasizing that the WMAP observations are more discrepant from our
X-ray predictions than by this factor, it must also be stressed that the ROSAT
$\beta$-model we employed are derived by fitting {\it only} the ROSAT data
from regions lying beyond `cooling flow' radii.
Thus, while it is true that
the central cooling of the hot ICM causes $\Delta T_{{\rm SZ}} (0)$
to decrease, any 
accompanying central 
peaking of the gas density which has not been taken into account by 
the $\beta$-model
has the opposite effect.  The net outcome depends
on how the product $n_e(r) kT(r)$, i.e. the gas
pressure, scales with radius {\it relative} to the
$\beta$-model inferred without taking the `cooling flow' into account.
After investigating our present sample, we found that `cooling flows'
tend to {\it maintain or raise} the $y_0$ parameter rather than lower it.  We
illustrate this point by showing in detail the situation of A2029, a
cluster over which the discrepancy between WMAP and ROSAT is large
(see Figure \ref{wmap}).  For this cluster the product $n_e(r) kT(r)$ is
constrained using the high resolution data of XMM-Newton, and its deprojected
radial profile as plotted in Figure \ref{xmm} is shown to compare closely with
the $\beta$-model of Table 1 for A2029.

{\it Could line-of-sight non-CMB emissions have contaminated the
WMAP passbands?}
In the
restricted venue of the center of the Coma cluster Herbig, Lawrence, \&
Readhead (1995) measured a SZE of -270 $\mu$K when the prediction
is $\approx$ -500 $\mu$K.  The authors attributed the discrepancy to
radio sources located along the central sightline but not members of
the cluster.
If, {\it in general}, line-of-sight sources unrelated
to the clusters are bright enough to affect the WMAP
data, the phenomenon should also
be present among non-cluster directions, i.e.
one should expect the
same level of contamination to exist in the WMAP random fields.  There
is however no evidence for this, because the radial profiles of the $\sim$ 30
accumulated random fields reveal an average temperature deviation of
$\approx$ 0.005 mK for the three cosmological bands of Q,V, and W
(see Figure \ref{random}), which
is on par with the average asymptotic
deviation among the co-added cluster fields.
Moreover, such amounts are far less than the discrepancy between
the predicted and observed SZE of Figure
\ref{sze-prediction}.

{\it Could the clusters themselves be a significant source of emission
in the WMAP passbands?}  There are two possibilities.  Diffuse emission
from cosmic ray synchrotron radiation and discrete radio sources.
The former is in principle a possibility (e.g. Sarazin \& Lieu 1998), and
we are currently investigating it, but see the two paragraphs after next.
Concerning the latter,
one can get a good idea of the cluster radio source occurence probability
by appealing to the Owens Valley radio interferometry survey (Bonamente et al
2005), which finds on  average
approximately one radio source per cluster with $\sim$ 1 mJy brightness at
30 GHz.  Since the Owens Valley sample is more distant than ours, with
a mean separation of $\sim$ 1.5 Gpc as opposed to $\sim$ 0.5 Gpc  for our
present sample, the brightness level  should
scale to 10 mJy, or 10$^{-28}$ W m$^{-2}$ Hz$^{-1}$.
Since, to reduce the SZE by invoking emission components, such
components must account for an average
CMB temperature increase 
of $\delta T_{{\rm CMB}} \approx$
5 $\times$ 10$^{-5}$ K distributed over the area of 0.5$^{\circ}$
angular radius.

Is the equivalent of one unresolved 10 mJy source within
the same area sufficient?  The conversion from $\delta T_{{\rm CMB}}$ to a
change in the observed flux involves multiplying the Rayleigh-Jeans sky flux
$2\pi k\delta T_{{\rm CMB}} \nu^2/c^2$   by the solid angle factor
$\delta\Omega/4\pi$ where $\delta\Omega = \pi \theta^2$ with
$\theta =$ 0.5$^{\circ}$.  This yields an excess flux at $\nu =$ 30 GHz
of 10$^{-27}$ W m$^{-2}$ Hz$^{-1}$, ten times higher than the
contribution from cluster radio sources.  Such a conclusion is applicable
to the Q band, which measures at a frequency only slightly higher than 30 GHz.
For the W band, at a frequency of 94 GHz, the radio source contribution
is at least two orders of magnitude short, because such sources typically
have flux spectra $F_{\nu} \sim \nu^{-\alpha}$ where $\alpha \geq$ 2.

Since any cluster emissions which may account for the less than
expected SZE detection by WMAP, be they diffuse or discrete in form,
are invariably non-thermal
in nature, with a flux spectrum $F_{\nu} \sim \nu^{-2}$ or
steeper,  i.e. a very different spectral shape from the Rayleigh-Jeans
$F_{\nu} \sim \nu^2$ dependence,
another test would involve checking the WMAP band ratios of
the SZE discrepancy.  Thus a cluster source is
distinguished by its larger V to W band flux ratios, which even
in the case of a power-law as shallow as
$F_{\nu} \sim \nu^{-0.5}$, is a factor of 
three more than the corresponding
ratio for Rayleigh-Jeans spectra.
Since the observed ratio of V:W for any extra radiation component
that may account for the discrepancy in Figure
\ref{sze-prediction}  is close to the black body limit, see
Figure \ref{ratio},
we conclude that the resolution does not lie with discrete cluster
radio sources.

In addition to the above, there is one more powerful test.  If
clusters do exhibit a normal SZE which is masked by self emissions
unrelated to the hot ICM, one would expect additional field-to-field
variation of the CMB temperature at a given radial bin within our
cluster sample, because the properties of the hot ICM and non-thermal
radiation vary from cluster to cluster.  At the very minimum, the
r.m.s. fluctuation must equal that of the blank field and the
differing degree of SZE from one hot ICM to the next, with the two
effects added in quadrature.  That even this minimum is already {\it less}
than the observed r.m.s. within our cluster sample is depicted in
Figure \ref{rms}.  In fact, the observed r.m.s. is at the same level as that
of the blank fields, excluding not only non-thermal emissions, but
also the SZE itself, unless one contrives a scenario in which the two
are correlated with each other.

{\it Could peculiarities in the hot ICM abundance, or the
gradual decline of the hot ICM temperature with radius be responsible?}
The general reasoning can always be applied, especially at the centers
of clusters where physics are more complicated, to argue that if there
are bright and spectrally unresolved metallic lines the density of the
hot ICM electrons, which is ordinarily determined by assuming that over
most the X-ray passband the emission is a continuum, may drop, resulting in
a reduction of the expected SZE.  In reality, however, the  effect is
very small.  Thus e.g. in the case of A2029, the hot ICM abundance originally
used by us to calculate the SZE was 0.3 solar.  Analysis of XMM-Newton data
revealed the same abundance level, Figure \ref{xmm}, hence no change in the
density of hot ICM electrons.  As to the large scale radial temperature
decline, numerical simulation of clusters (Romeo et al
2005) indicates that the 
hot ICM temperature can be halved by the time one reaches the virial radius.
Observationally, however, it is clear that at least out to the radius
$R = 5r_c$
beyond which the hot ICM contribution of the SZE is negligible (see Eq. (7)),
there is no evidence for such a temperature drop.  For instance, in the
case of A2029 which has a core radius of 2 arcmin, the hot ICM temperature
as measured by XMM-Newton (Figure \ref{xmm})
remains stable out to 10 arcmin.  Thus the
cooling of the hot ICM towards the cluster's outskirts is not the reason
for the large discrepancy between WMAP and X-ray fluxes.

{\it On the role of asphericity of either the central core or the entire
cluster, and clumpiness of the hot ICM}, these questions are harder to
answer quantitatively.  Given that we averaged over 31 clusters, however,
any residual asphericity 
corections must be very small, unless one contrives a scenario
under which clusters have their major axes aligned along some preferred
direction.  Clumpiness is a phenomenon worthy of further research,
because (to argue heuristically)
while the X-ray observations measure $\langle n^2_e \rangle$
the SZE probes $\langle n_e \rangle$.  
The two are trivially related to each other if $\langle n_e \rangle^2 =
\langle n^2_e \rangle$,
which occurs only when the gas is smooth.  Once
clumps are present, $\langle n^2_e \rangle > \langle n_e \rangle^2$,
so the X-ray prediction will
overestimate the SZE flux.
Although for the hot ICM beyond the central core it is difficult to
envisage, from pressure balance considerations, how such high temperature
plasmas could exhibit a large degree of clumping, the situation may
be different within the core itself where the cooling of gas
and other dynamic processes often occur.  It would still be surprising
if the factor of six discrepancy reported here for the
W-band could entirely be attributed to this effect. Nevertheless,
in the absence of further observational evidence, or a sound physical argument,
one should not exclude the scenario of hot ICM clumping as possible
cause of our reported discrepancy.

Finally, one could ask {\it if photon populations other than
the CMB may exist to interact with the hot ICM electrons to
`refill' the SZE flux}.
Whether we are dealing with `upward' or `downward' scattering,
the difficulty lies with insufficient seed photon flux
at frequencies well below and above the WMAP passbands.  This
is the reason for the general
belief that cluster SZE should cause a clean removal
of the CMB from its original microwave passband of emission.

\vspace{2mm}

\noindent
{\bf Conclusion}

Formally, a statistically significant detection of the SZE 
across our entire cluster sample was achieved
by WMAP.  The level of this detection is very weak, however.
Not only is
the field-to-field variation of the CMB temperature
within the sample (due to different clusters exerting
unequal SZE on the CMB) non-existent, Figure \ref{rms}, but also
the {\it measured} decrements of Figure
\ref{sze-prediction} are
consistent with nothing beyond the usual primary CMB anisotropy, viz.
the systematic variations in the
CMB average radial profile for a commensurate number of co-added 
blank (random) fields,
Figure \ref{random}.  Of particular concern are the data of the W-band, which
is the passband containing the cleanest extragalactic signals (Bennett et al
2003b).  The statistical significance of an overall W-band SZE detection is $<$
3 $\sigma$, when the expected
effect is far larger.  Here we also note that a similar shallow decrement was
seen in the Myers et al. (2004) paper where the best fit model had a $\Delta
T_0$ of 0.083 mK in the W band.  
This is much smaller than the predicted average
decrement from our sample of 0.46 mK from Table 1.  
Thus, taken at
value one may even hold the opinion 
that there is in fact no strong evidence in the WMAP database for
the SZE at all,
{\it when the aggregate behavior of all
the clusters in the sample}, rather than individual cases, is considered.  

Naturally the entire premise of this paper depends
on the reliability of the original WMAP data.  If
there are any data analysis issues with the WMAP processing that can explain
the extra diffuse emission seen in our SZE clusters then our findings
will be obsolete.
However, this would have implicate severely all the WMAP analysis done to
date.  One possible resolution is to look at the SZE as
probed using dedicated ground based observatories.  The SZE has already been
detected in a large number of high-redshift clusters using interferometric
techniques of higher resolution than the WMAP data (e.g., Carlstrom et
al. 1996; Joy et al. 2001; Reese et al. 2002; Bonamente et al. 2005; LaRoque et
al. 2005).  Comparison of radio interferometry and X-ray data for the same
clusters show that SZE-derived and X-ray derived masses and gas fractions are
in agreement (Grego et al. 2001; LaRoque et al. 2005), and allows for a
determination of the cosmic distance scale (Reese et al. 2002;
Bonamente et al. 2005).  There is not necessarily a conflict between our
present results and the previous reported SZE detections for individual
clusters.  As can be seen from Figure \ref{wmap}, many of the clusters
in our sample {\it do} exhibit the effect at $\sim$ the anticipated level.

In summary, it is through the first detailed {\it radial
profile} comparison
between X-ray and microwave observations that an apparent sample-wide
discrepancy between the expected and measured levels of SZE from some of the
best known clusters of galaxies was uncovered.
The difficulty lies with the
average behavior of our randomly selected cluster sample, which could 
still be suffering from
systematic yet hitherto unknown biases.  Nonetheless,
the average CMB temperature decrement is sufficiently shallow to
be interpreted simply as the usual primary CMB anisotropy: there
is no need to invoke any SZE at this stage of the WMAP analysis.

We are grateful to an anoymous referee for
comments and critique of this work.

\begin{deluxetable}{lrrccccccccc}
\tablewidth{19cm}
\tabletypesize{\scriptsize}
\rotate
 
\tablecaption{X-ray properties
of the 31 clusters employed for the present SZE investigation.  The
$\beta$-model parameters are derived from ROSAT observations.
 \label{table:beta}}
\renewcommand{\arraystretch}{0.70}
\startdata
Name & \multicolumn{2}{c}{Galactic (2000)} & Redshift &kT & n$_0$ & $\beta$ & R$_{core}$ & $\Delta T0_Q$ & $\Delta T0_V$ &$\Delta T0_W$ & CF?\\ 
     & Long. & Lat. & & (keV) & $\times 10^{-3}$ cm$^{-3}$ & & (arcmin) & mK & mK & mK & \\ \hline
Abell 85   &  115.053 & -72.064 & 0.055   & 7.0 & $15.17^{+0.30}_{-0.40}$ & $0.58^{+0.03}_{-0.04}$ & $   1.8^{  +0.5}_{  -0.9}$ & $-0.99_{-0.30}^{+0.49}$ & $-0.94_{-0.28}^{+0.46}$ & $-0.81_{-0.24}^{+0.40}$ & Yes \\ 
Abell 133  &  149.761 & -84.233 & 0.057   & 5.0 & $ 3.33^{+0.14}_{-0.13}$ & $0.72^{+0.09}_{-0.07}$ & $   3.4^{  +0.8}_{  -0.8}$ & $-0.22_{-0.06}^{+0.06}$ & $-0.20_{-0.06}^{+0.06}$ & $-0.18_{-0.05}^{+0.05}$ & Yes\\ 
Abell 665  &  149.735 &  34.673 & 0.1816  & 7.0 & $ 3.34^{+0.19}_{-0.23}$ & $0.64^{+0.10}_{-0.10}$ & $   1.3^{  +0.1}_{  -0.1}$ & $-0.40_{-0.14}^{+0.08}$ & $-0.38_{-0.14}^{+0.08}$ & $-0.33_{-0.12}^{+0.07}$ & No\\ 
Abell 1068 &  179.100 &  60.130 & 0.139   & 5.0 & $ 8.16^{+0.49}_{-0.43}$ & $0.90^{+0.17}_{-0.13}$ & $   1.5^{  +0.5}_{  -0.5}$ & $-0.43_{-0.16}^{+0.15}$ & $-0.40_{-0.15}^{+0.15}$ & $-0.35_{-0.13}^{+0.13}$ & Yes\\ 
Abell 1302 &  134.668 &  48.904 & 0.116   & 4.8 & $ 2.88^{+0.19}_{-0.15}$ & $0.64^{+0.12}_{-0.08}$ & $   1.4^{  +0.4}_{  -0.3}$ & $-0.17_{-0.07}^{+0.05}$ & $-0.16_{-0.06}^{+0.05}$ & $-0.14_{-0.05}^{+0.04}$ & \nodata\\ 
Abell 1314 &  151.828 &  63.567 & 0.0341  & 5.0 & $ 1.00^{+0.27}_{-0.26}$ & $0.35^{+0.21}_{-0.10}$ & $   2.6^{  +3.2}_{  -2.4}$ & $-0.44_{ nan}^{+0.58}$ & $-0.42_{ nan}^{+0.55}$ & $-0.36_{ nan}^{+0.47}$ & No\\ 
Abell 1361 &  153.292 &  66.581 & 0.1167  & 4.0 & $ 2.81^{+nan}_{-0.73}$ & $1.78^{+18.23}_{-1.08}$ & $   5.2^{ +34.0}_{  -4.1}$ & $-0.19_{ nan}^{+0.22}$ & $-0.18_{ nan}^{+0.21}$ & $-0.16_{ nan}^{+0.18}$ & Yes\\ 
Abell 1367 &  234.799 &  73.030 & 0.0276  & 3.5 & $ 1.68^{+0.03}_{-0.03}$ & $0.52^{+0.02}_{-0.02}$ & $   8.6^{  +0.5}_{  -0.6}$ & $-0.17_{-0.02}^{+0.02}$ & $-0.16_{-0.02}^{+0.02}$ & $-0.14_{-0.02}^{+0.01}$ & uncertain\\ 
Abell 1413 &  226.182 &  76.787 & 0.143   & 6.0 & $13.85^{+0.78}_{-0.94}$ & $0.68^{+0.11}_{-0.11}$ & $   1.1^{  +0.1}_{  -0.1}$ & $-0.91_{-0.29}^{+0.18}$ & $-0.86_{-0.28}^{+0.17}$ & $-0.74_{-0.24}^{+0.14}$ & No\\ 
Abell 1689 &  313.387 &  61.097 & 0.181   & 7.0 & $14.02^{+0.76}_{-0.90}$ & $0.75^{+0.12}_{-0.12}$ & $   1.0^{  +0.0}_{  -0.0}$ & $-1.01_{-0.28}^{+0.17}$ & $-0.96_{-0.26}^{+0.16}$ & $-0.83_{-0.23}^{+0.14}$ & No\\ 
Abell 1795 &  33.788  &  77.155 & 0.061   & 7.0 & $ 3.05^{+0.04}_{-0.06}$ & $0.99^{+0.04}_{-0.06}$ & $   5.2^{  +0.3}_{  -0.4}$ & $-0.33_{-0.03}^{+0.03}$ & $-0.32_{-0.03}^{+0.03}$ & $-0.27_{-0.02}^{+0.02}$ & Yes\\ 
Abell 1914 &  67.196  &  67.453 & 0.171   & 9.0 & $13.34^{+0.23}_{-0.21}$ & $0.85^{+0.04}_{-0.04}$ & $   1.4^{  +0.1}_{  -0.1}$ & $-1.47_{-0.12}^{+0.12}$ & $-1.39_{-0.12}^{+0.11}$ & $-1.20_{-0.10}^{+0.10}$ & \nodata\\ 
Abell 1991 &  22.762  &  60.497 & 0.0586  & 4.0 & $ 3.18^{+0.56}_{-0.35}$ & $0.82^{+0.54}_{-0.22}$ & $   2.8^{  +2.8}_{  -8.4}$ & $-0.12_{-0.14}^{+0.37}$ & $-0.12_{-0.13}^{+0.35}$ & $-0.10_{-0.11}^{+0.30}$ & Yes\\ 
Abell 2029 &  6.505   &  50.547 & 0.0767  & 9.0 & $13.54^{+0.21}_{-0.92}$ & $0.67^{+0.03}_{-0.11}$ & $   1.9^{  +0.3}_{  -0.3}$ & $-1.27_{-0.44}^{+0.23}$ & $-1.21_{-0.42}^{+0.21}$ & $-1.04_{-0.36}^{+0.18}$ & Yes\\ 
Abell 2142 &  44.213  &  48.701 & 0.09    & 9.0 & $ 7.09^{+0.40}_{-0.48}$ & $0.68^{+0.11}_{-0.11}$ & $   2.4^{  +0.2}_{  -0.3}$ & $-0.98_{-0.32}^{+0.21}$ & $-0.93_{-0.30}^{+0.20}$ & $-0.80_{-0.26}^{+0.18}$ & Yes\\ 
Abell 2199 &  62.897  &  43.697 & 0.0302  & 4.5 & $ 6.87^{+0.08}_{-0.23}$ & $0.65^{+0.02}_{-0.05}$ & $   3.1^{  +0.5}_{  -2.5}$ & $-0.23_{-0.05}^{+0.19}$ & $-0.22_{-0.05}^{+0.18}$ & $-0.19_{-0.04}^{+0.16}$ & Yes\\ 
Abell 2218 &  97.745  &  38.124 & 0.171   & 6.0 & $ 3.00^{+0.17}_{-0.20}$ & $0.72^{+0.12}_{-0.12}$ & $   1.5^{  +0.1}_{  -0.1}$ & $-0.27_{-0.08}^{+0.05}$ & $-0.26_{-0.08}^{+0.05}$ & $-0.22_{-0.07}^{+0.04}$ & uncertain\\ 
Abell 2219 &  72.597  &  41.472 & 0.228   & 7.0 & $ 6.12^{+0.14}_{-0.13}$ & $0.78^{+0.05}_{-0.04}$ & $   1.8^{  +0.2}_{  -0.1}$ & $-0.88_{-0.10}^{+0.09}$ & $-0.83_{-0.10}^{+0.09}$ & $-0.72_{-0.08}^{+0.08}$ & No\\ 
Abell 2241 &  54.784  &  36.643 & 0.0635  & 3.1 & $11.66^{+0.48}_{-0.43}$ & $0.74^{+0.09}_{-0.07}$ & $   1.0^{  +0.2}_{  -0.2}$ & $-0.15_{-0.03}^{+0.03}$ & $-0.14_{-0.03}^{+0.03}$ & $-0.12_{-0.03}^{+0.03}$ & \nodata \\ 
Abell 2244 &  56.772  &  36.306 & 0.097   & 7.0 & $30.50^{+0.62}_{-1.00}$ & $0.59^{+0.03}_{-0.04}$ & $   1.0^{  +0.4}_{  -0.9}$ & $-1.72_{-0.71}^{+1.67}$ & $-1.63_{-0.67}^{+1.58}$ & $-1.41_{-0.58}^{+1.36}$ & Yes\\ 
Abell 2255 &  93.975  &  34.948 & 0.08    & 7.0 & $ 2.49^{+0.05}_{-0.05}$ & $0.76^{+0.04}_{-0.04}$ & $   4.6^{  +0.4}_{  -0.3}$ & $-0.40_{-0.04}^{+0.04}$ & $-0.38_{-0.04}^{+0.04}$ & $-0.33_{-0.03}^{+0.03}$ & No\\ 
Abell 2256 &  111.096 &  31.738 & 0.06    & 7.0 & $ 3.74^{+0.19}_{-0.23}$ & $0.85^{+0.14}_{-0.14}$ & $   5.5^{  +0.9}_{  -0.9}$ & $-0.48_{-0.14}^{+0.11}$ & $-0.46_{-0.13}^{+0.10}$ & $-0.39_{-0.11}^{+0.09}$ & No\\ 
Abell 2597 &  65.363  & -64.836 & 0.085   & 4.0 & $18.94^{+0.79}_{-0.71}$ & $0.69^{+0.08}_{-0.06}$ & $   1.2^{  +0.7}_{  -3.0}$ & $-0.52_{-0.33}^{+1.37}$ & $-0.49_{-0.31}^{+1.29}$ & $-0.43_{-0.27}^{+1.12}$ & Yes\\ 
Abell 2670 &  81.318  & -68.516 & 0.076   & 3.0 & $ 4.21^{+0.17}_{-0.18}$ & $0.64^{+0.07}_{-0.06}$ & $   1.9^{  +0.6}_{  -0.8}$ & $-0.14_{-0.05}^{+0.06}$ & $-0.13_{-0.05}^{+0.06}$ & $-0.12_{-0.04}^{+0.05}$ & Yes\\ 
Abell 2717 &  349.076 & -76.390 & 0.049   & 3.0 & $ 9.41^{+0.21}_{-0.19}$ & $0.64^{+0.04}_{-0.03}$ & $   1.5^{  +0.2}_{  -0.1}$ & $-0.17_{-0.02}^{+0.02}$ & $-0.16_{-0.02}^{+0.02}$ & $-0.14_{-0.02}^{+0.02}$ & \nodata\\ 
Abell 2744 &  8.898   & -81.241 & 0.308   &11.0 & $ 3.12^{+0.23}_{-0.17}$ & $1.60^{+0.44}_{-0.27}$ & $   3.3^{  +0.6}_{  -0.4}$ & $-0.87_{-0.22}^{+0.19}$ & $-0.82_{-0.21}^{+0.18}$ & $-0.71_{-0.18}^{+0.15}$ & No\\ 
Abell 3301 &  242.415 & -37.409 & 0.054   & 7.0 & $ 4.33^{+0.20}_{-0.18}$ & $0.49^{+0.05}_{-0.04}$ & $   1.8^{  +0.5}_{  -0.4}$ & $-0.39_{-0.15}^{+0.11}$ & $-0.37_{-0.14}^{+0.11}$ & $-0.32_{-0.12}^{+0.09}$ & \nodata\\ 
Abell 3558 &  311.978 &  30.738 & 0.048   & 5.0 & $ 2.65^{+0.04}_{-0.04}$ & $0.79^{+0.04}_{-0.03}$ & $   5.9^{  +0.4}_{  -0.4}$ & $-0.23_{-0.02}^{+0.02}$ & $-0.22_{-0.02}^{+0.02}$ & $-0.19_{-0.02}^{+0.02}$ & Yes\\ 
Abell 3560 &  312.578 &  28.890 & 0.04    & 2.0 & $ 4.96^{+0.23}_{-0.21}$ & $0.49^{+0.05}_{-0.04}$ & $   2.6^{  +0.6}_{  -0.5}$ & $-0.13_{-0.05}^{+0.04}$ & $-0.13_{-0.05}^{+0.04}$ & $-0.11_{-0.04}^{+0.03}$ & \nodata\\ 
Abell 3562 &  313.308 &  30.349 & 0.04    & 4.5 & $ 7.33^{+0.51}_{-0.64}$ & $0.47^{+0.08}_{-0.08}$ & $   1.3^{  +0.1}_{  -0.1}$ & $-0.26_{-0.26}^{+0.08}$ & $-0.25_{-0.25}^{+0.08}$ & $-0.22_{-0.21}^{+0.07}$ & No\\ 
Abell 3571 &  316.317 &  28.545 & 0.04    & 7.0 & $13.02^{+0.32}_{-0.30}$ & $0.65^{+0.04}_{-0.04}$ & $   3.6^{  +0.7}_{  -0.7}$ & $-1.05_{-0.23}^{+0.23}$ & $-1.00_{-0.22}^{+0.22}$ & $-0.86_{-0.19}^{+0.19}$ & Yes\\ 
Abell 4059 &  356.833 & -76.061 & 0.046   & 4.5 & $ 5.03^{+0.45}_{-0.34}$ & $0.99^{+0.30}_{-0.18}$ & $   6.0^{  +2.2}_{  -1.7}$ & $-0.31_{-0.13}^{+0.11}$ & $-0.29_{-0.13}^{+0.10}$ & $-0.25_{-0.11}^{+0.09}$ & Yes\\ 
Coma       &  58.080  &  87.958 & 0.023   & 8.2 & $ 4.42^{+0.24}_{-0.29}$ & $0.71^{+0.11}_{-0.11}$ & $   9.8^{  +1.6}_{  -1.6}$ & $-0.59_{-0.20}^{+0.14}$ & $-0.56_{-0.19}^{+0.13}$ & $-0.48_{-0.16}^{+0.11}$ & No\\ \hline

\enddata
\end{deluxetable}

\begin{figure}[H]
\begin{center}
\includegraphics[angle=0,width=6in]{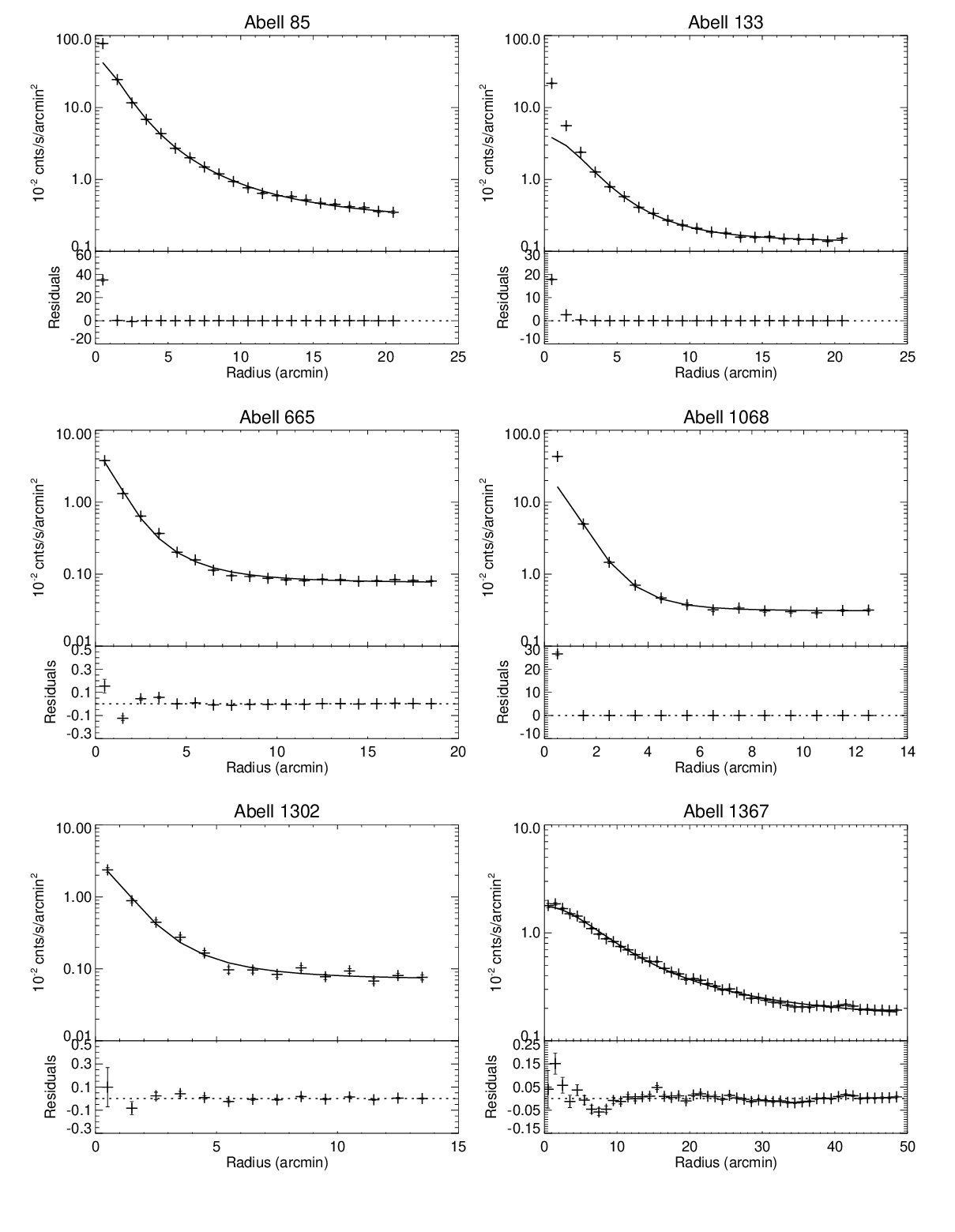}
\vspace{0.0mm}
\end{center}
\vspace{0.0cm}
\end{figure}

\begin{figure}[H]
\begin{center}
\includegraphics[angle=0,width=6in]{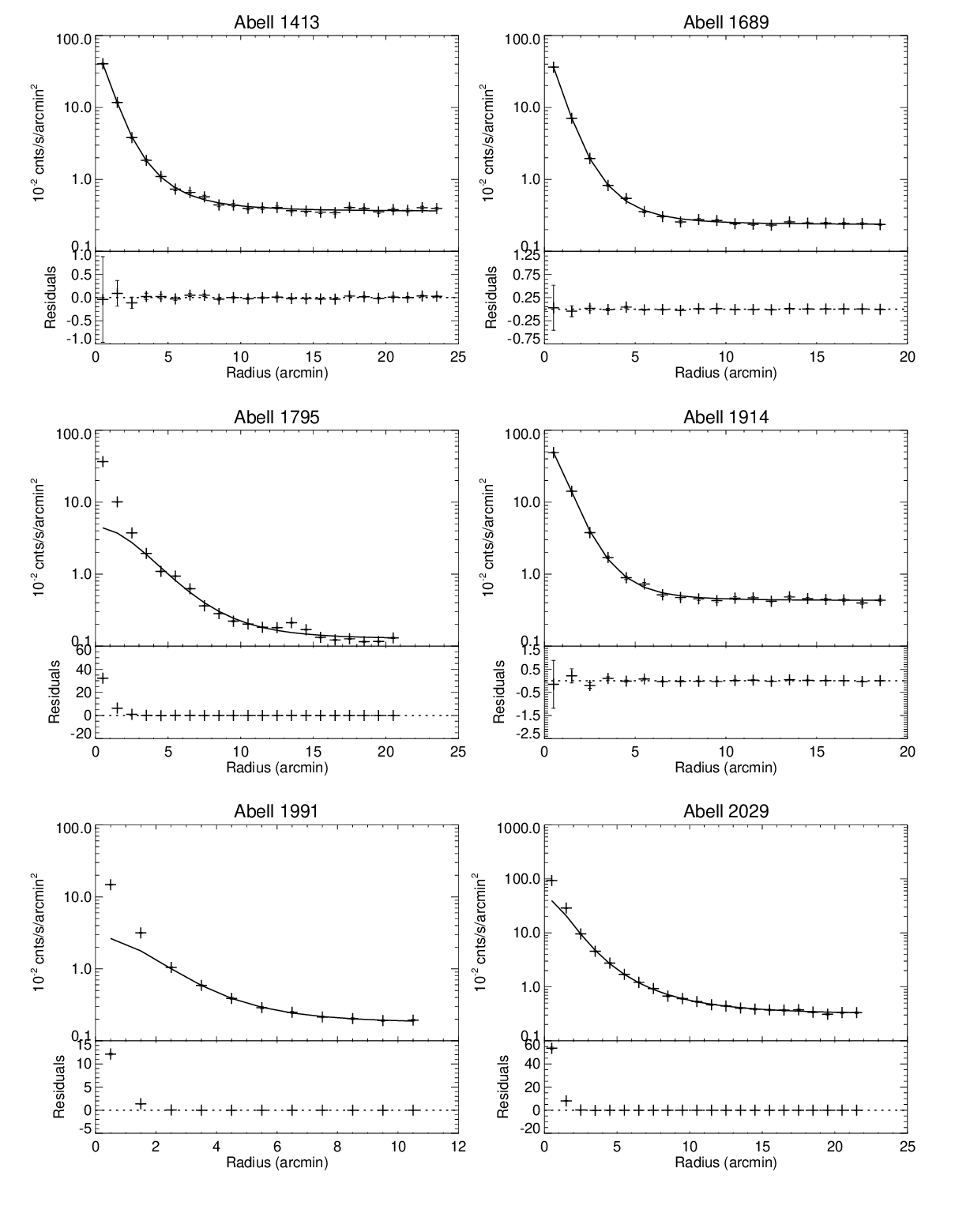}
\vspace{0.0mm}
\end{center}
\vspace{0.0cm}
\end{figure}

\begin{figure}[H]
\begin{center}
\includegraphics[angle=0,width=6in]{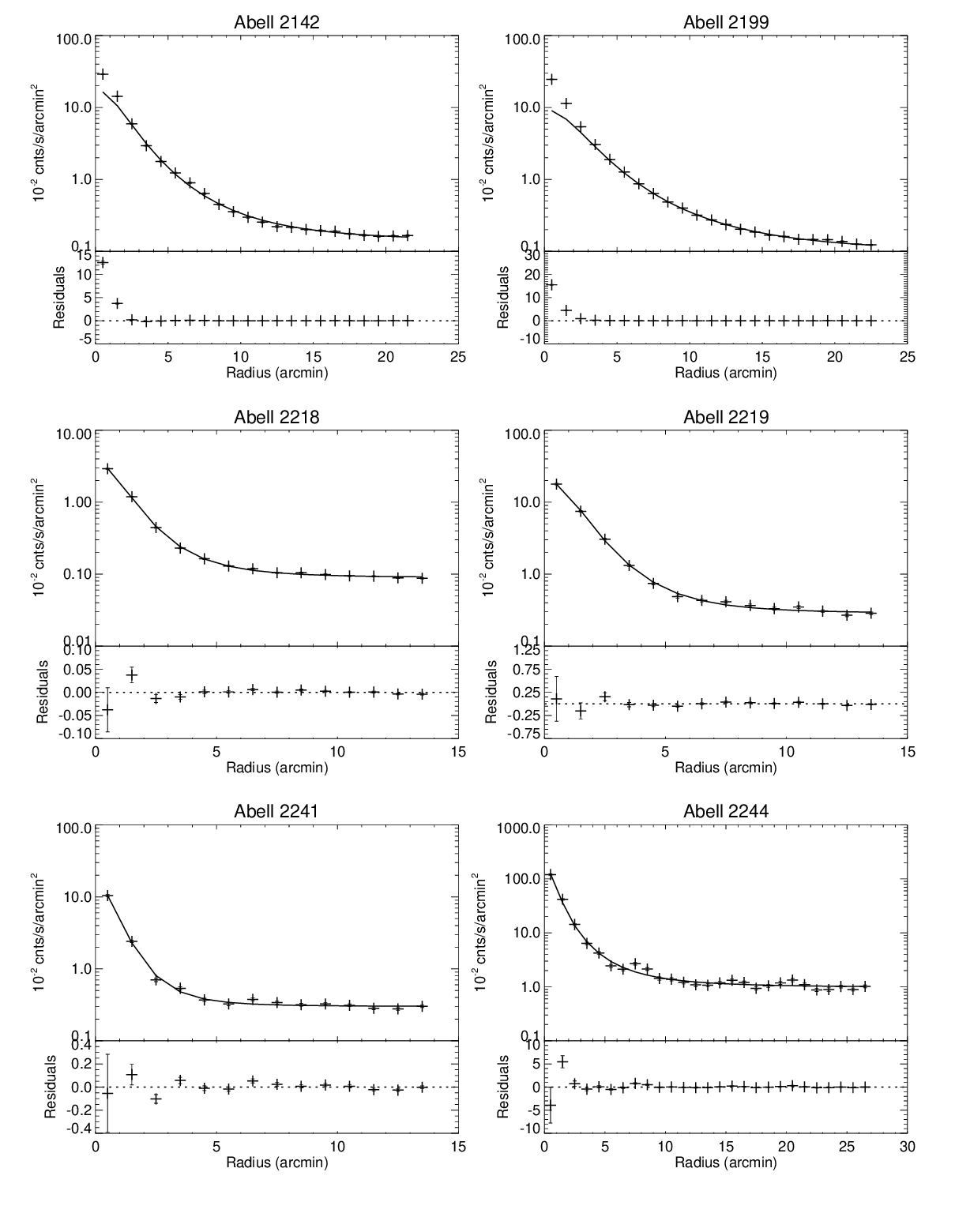}
\vspace{0.0mm}
\end{center}
\vspace{0.0cm}
\end{figure}
                                                                                          
\begin{figure}[H]
\begin{center}
\includegraphics[angle=0,width=6in]{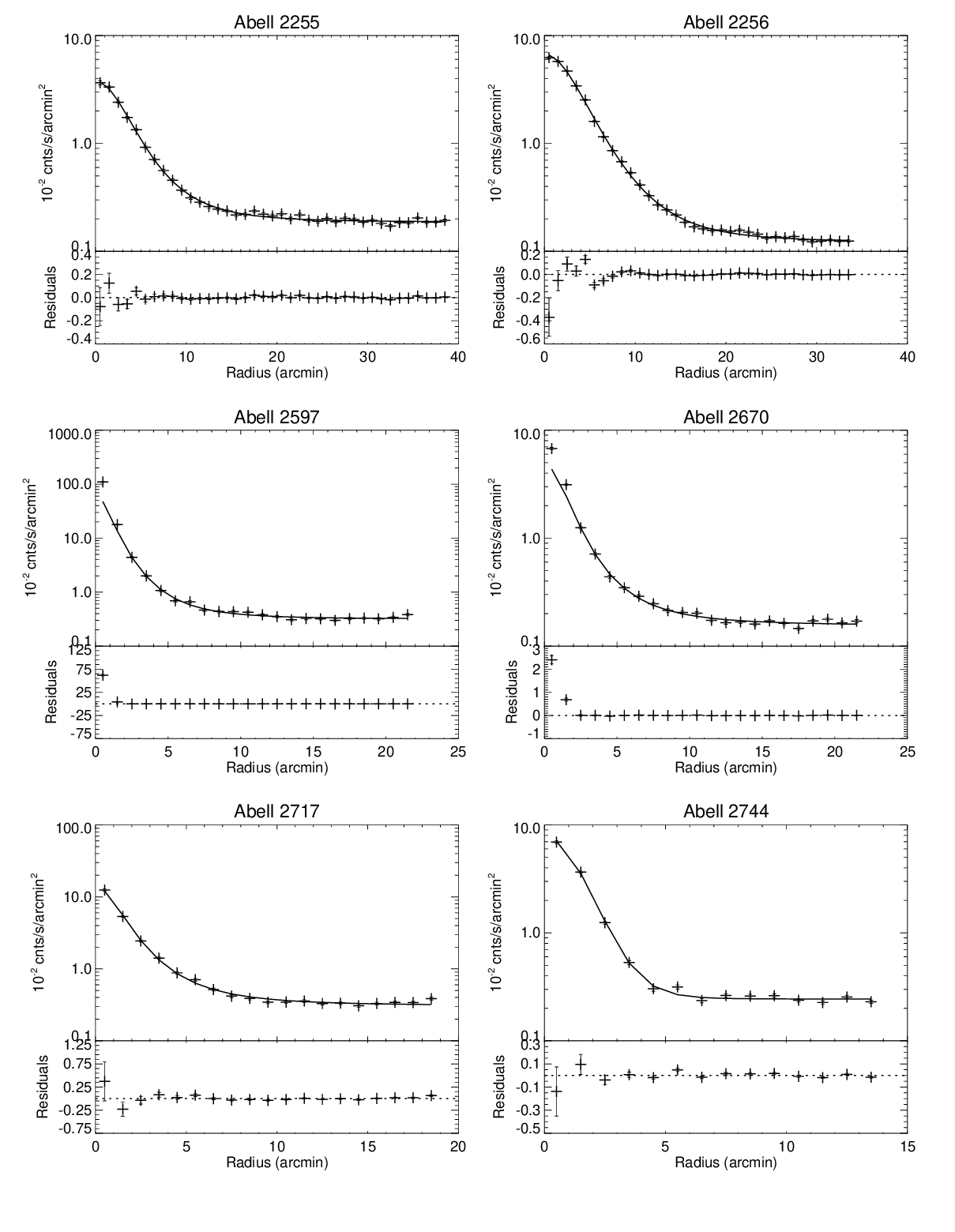}
\vspace{0.0mm}
\end{center}
\vspace{0.0cm}
\end{figure}

\begin{figure}[H]
\begin{center}
\includegraphics[angle=0,width=6in]{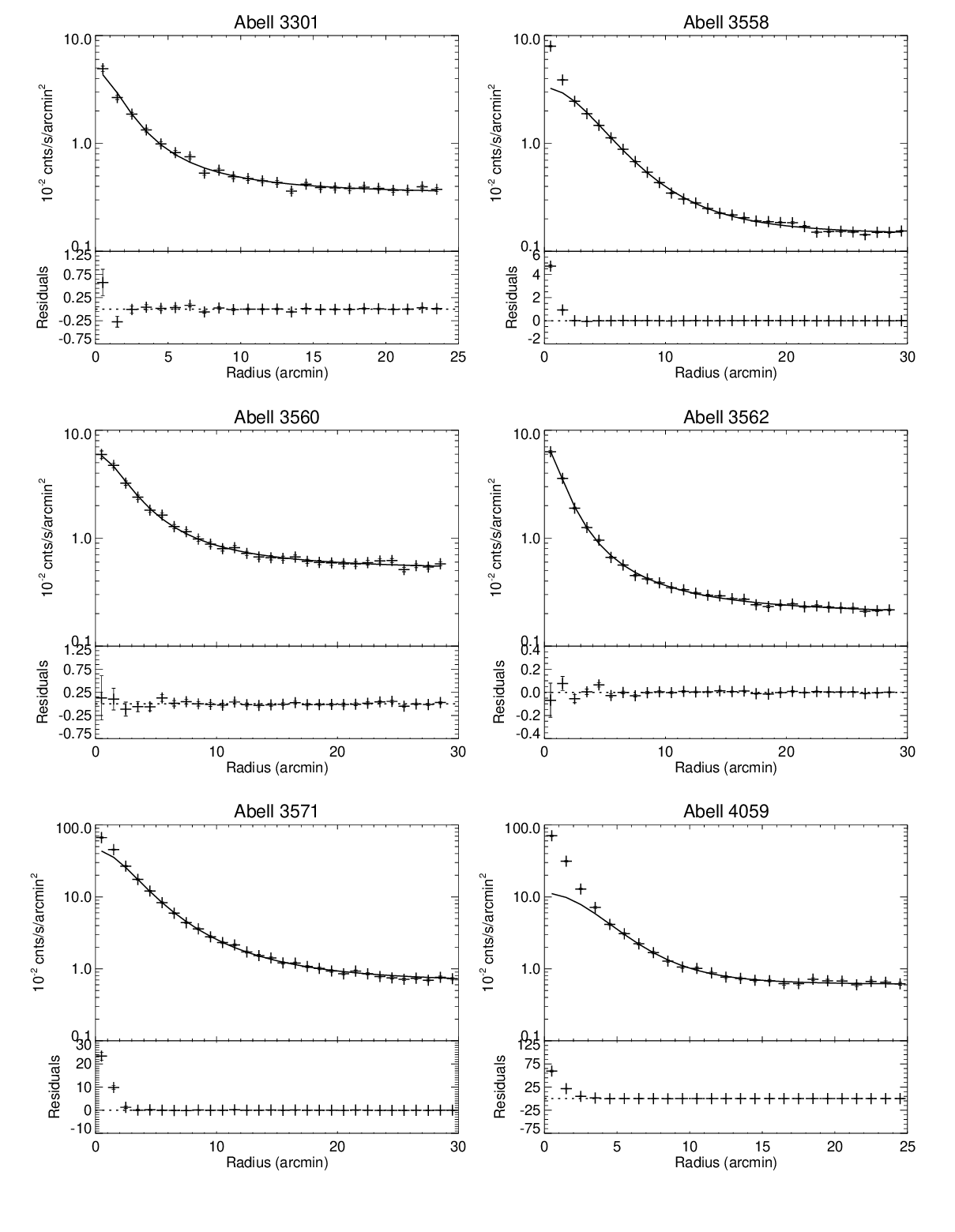}
\vspace{0.0mm}
\end{center}
\vspace{0.0cm}
\end{figure}

\begin{figure}[H]
\begin{center}
\includegraphics[angle=0,width=6in]{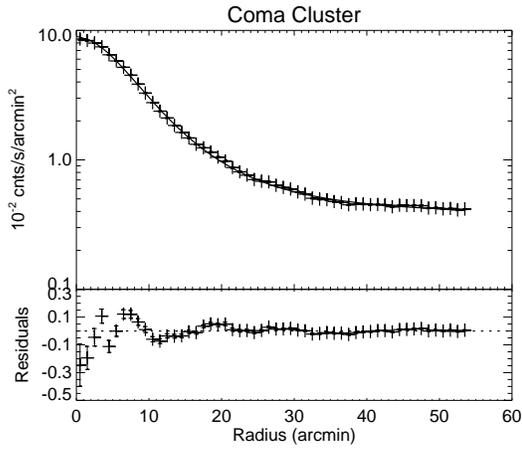}
\vspace{-10cm}
\end{center}
\caption{Isothermal $\beta$-model (Eq. 1)
fits to the ROSAT X-ray surface brightness
data of the Bonamente et al sample of clusters.  The ROSAT
mission was chosen because of its wide field-of-view, which
allows one to determine clearly the surrounding background
level even for the larger clusters.  For  non-isothermal
`cooling flow' clusters
the central region where the phenomenon occurs is excluded
from our analysis.
\label{rosat}}
\vspace{0.0cm}
\end{figure}

\begin{figure}[H]
\begin{center}
\includegraphics[angle=0,width=6in]{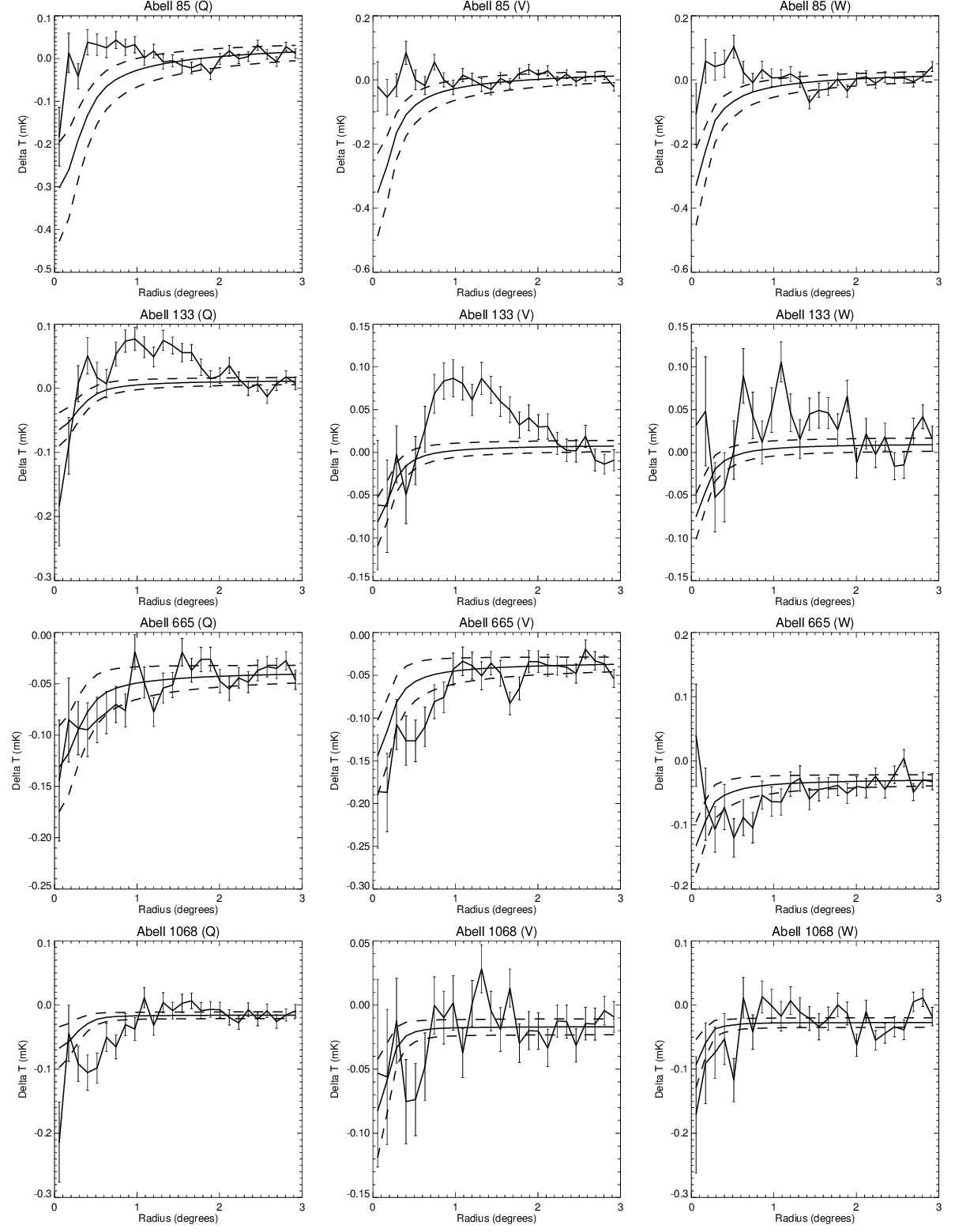}
\vspace{0.0mm}
\end{center}
\vspace{0.0cm}
\end{figure}

\begin{figure}[H]
\begin{center}
\includegraphics[angle=0,width=6in]{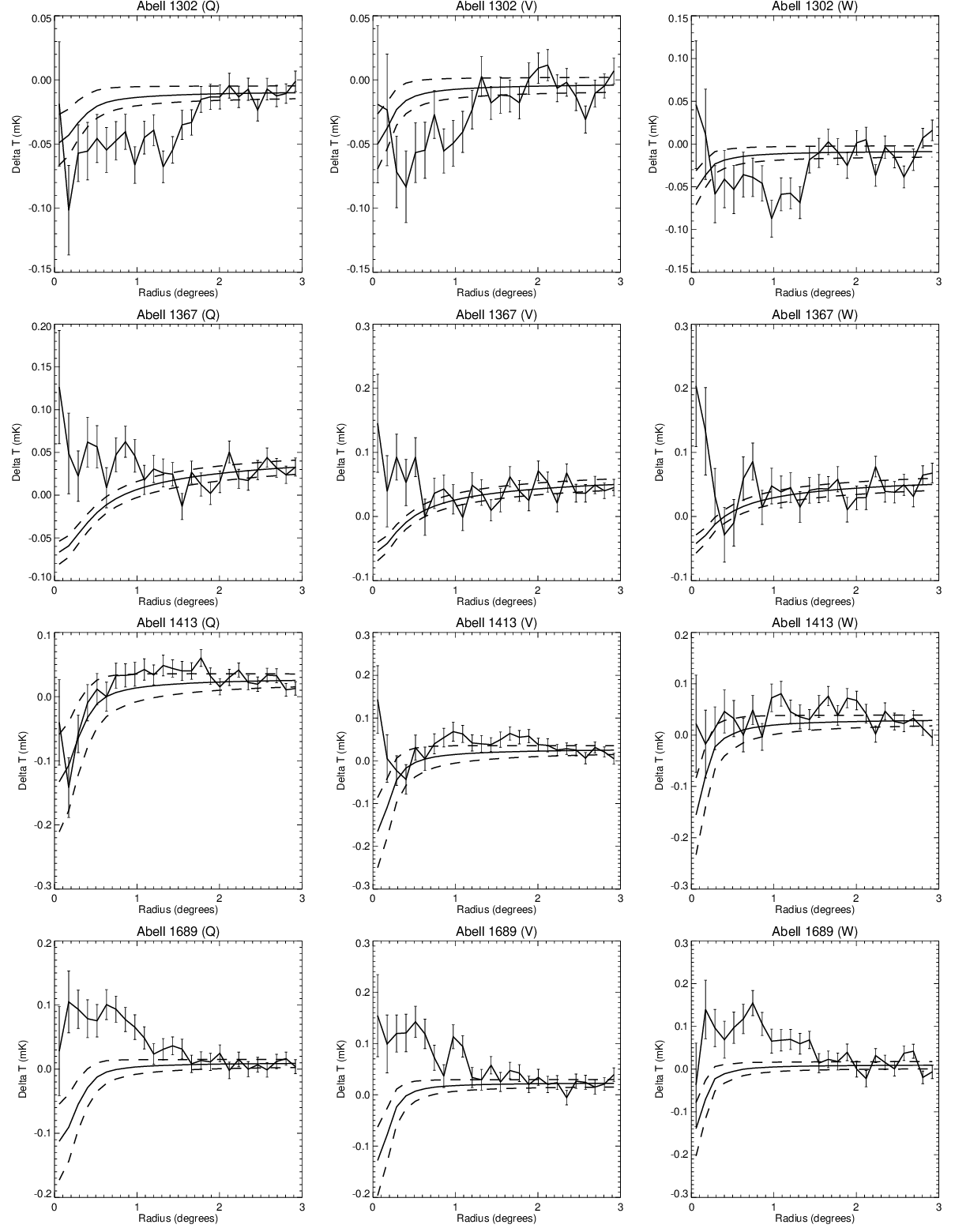}
\vspace{0.0mm}
\end{center}
\vspace{0.0cm}
\end{figure}

\begin{figure}[H]
\begin{center}
\includegraphics[angle=0,width=6in]{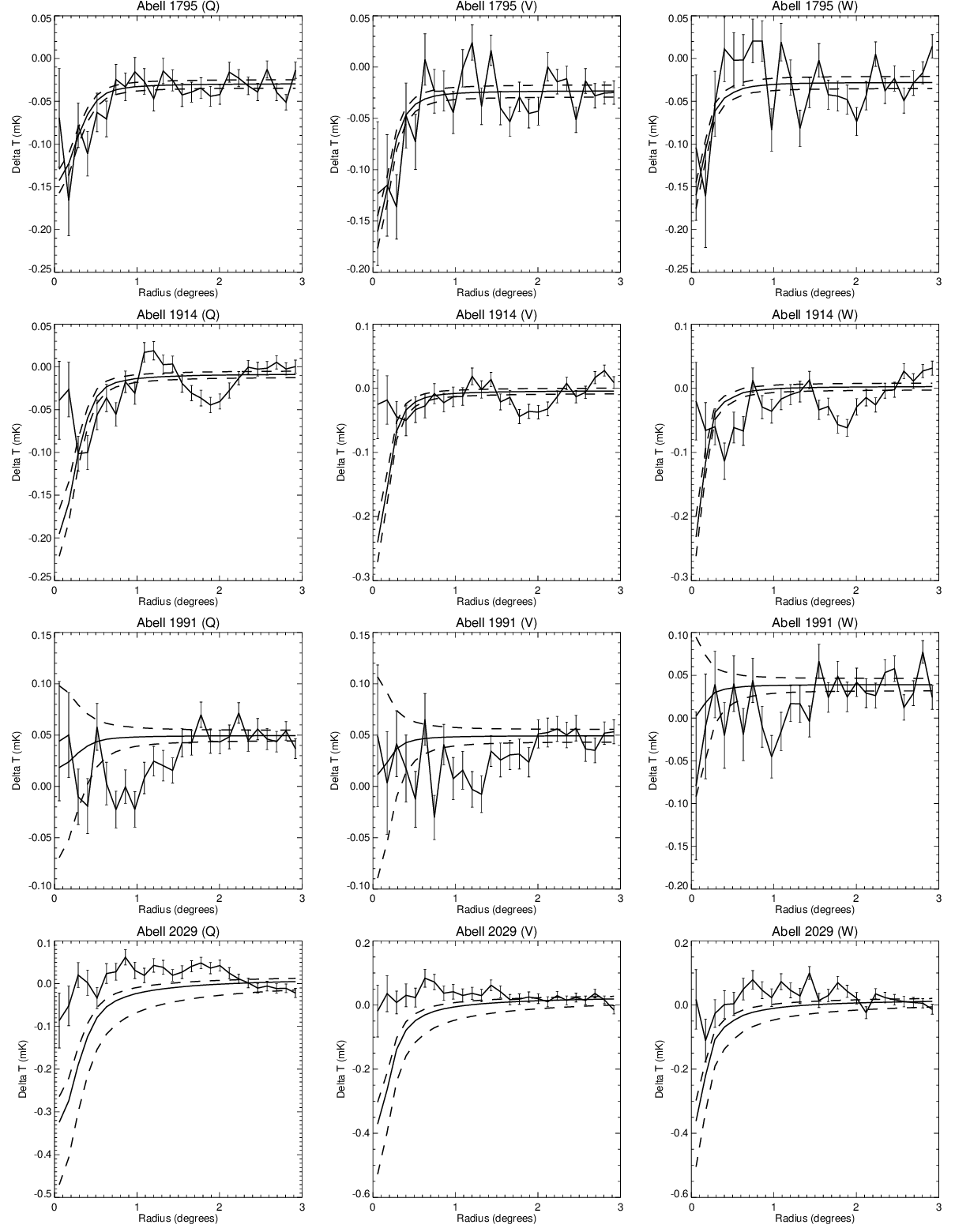}
\vspace{0.0mm}
\end{center}
\vspace{0.0cm}
\end{figure}

\begin{figure}[H]
\begin{center}
\includegraphics[angle=0,width=6in]{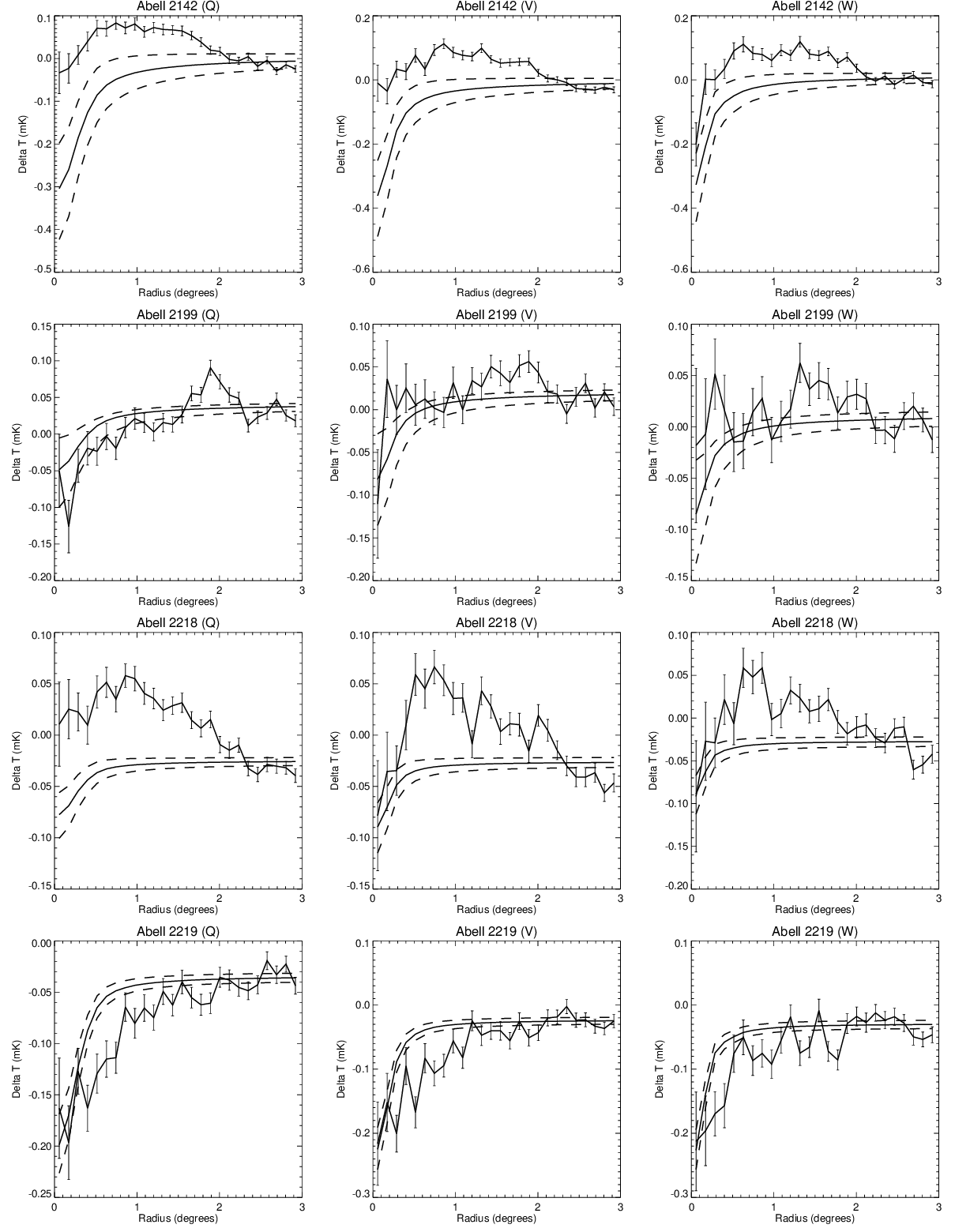}
\vspace{0.0mm}
\end{center}
\vspace{0.0cm}
\end{figure}

\begin{figure}[H]
\begin{center}
\includegraphics[angle=0,width=6in]{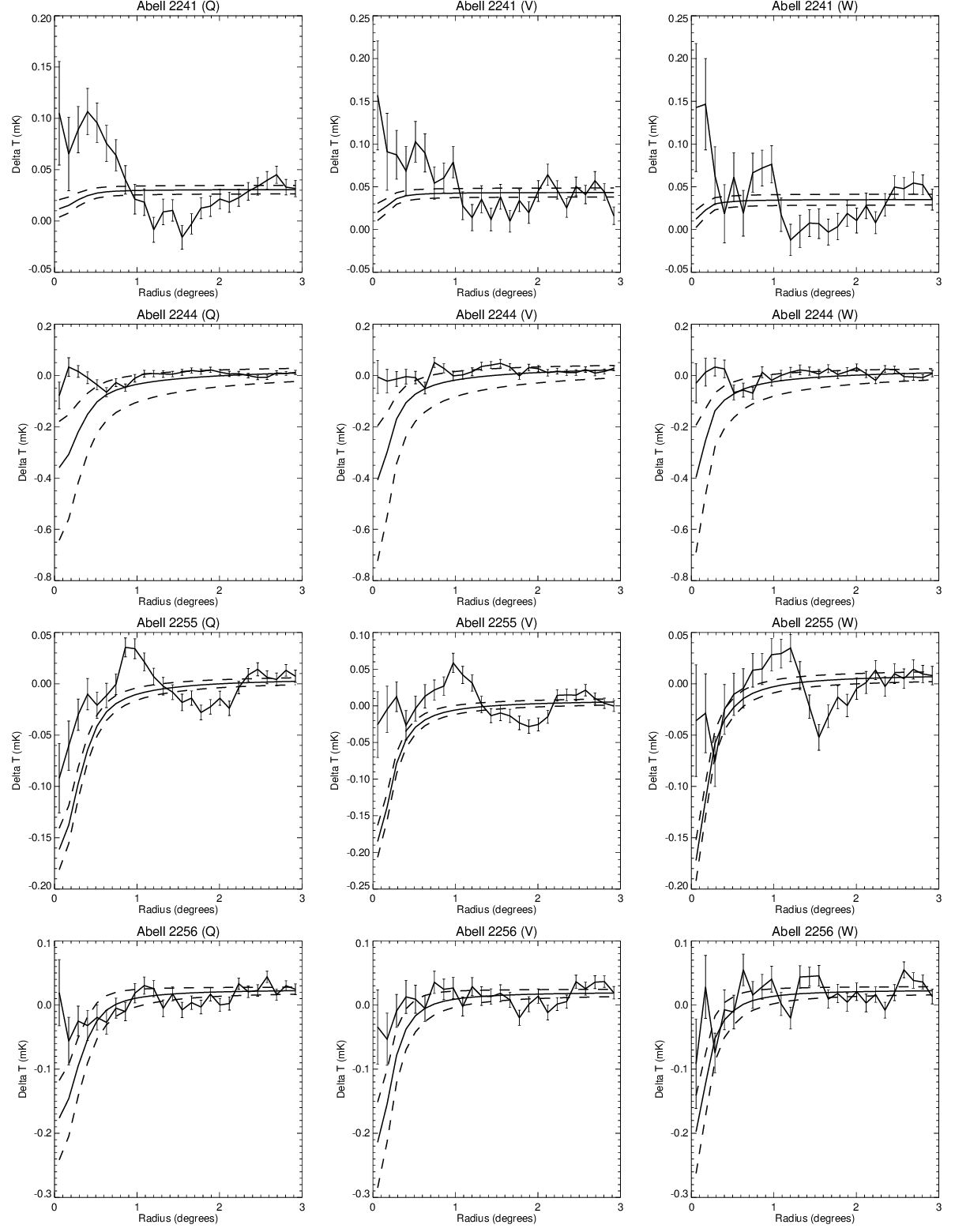}
\vspace{0.0mm}
\end{center}
\vspace{0.0cm}
\end{figure}

\begin{figure}[H]
\begin{center}
\includegraphics[angle=0,width=6in]{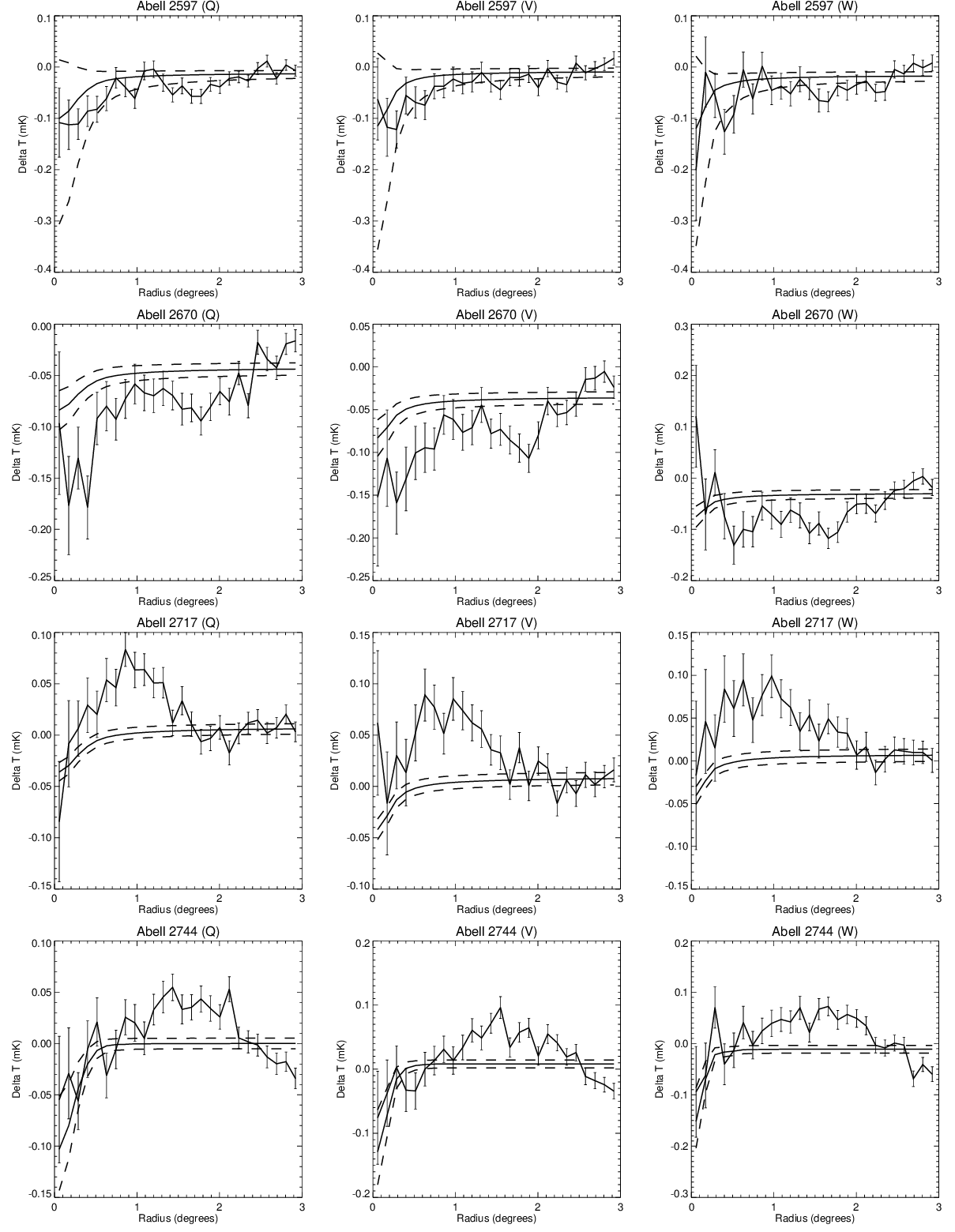}
\vspace{0.0mm}
\end{center}
\vspace{0.0cm}
\end{figure}

\begin{figure}[H]
\begin{center}
\includegraphics[angle=0,width=6in]{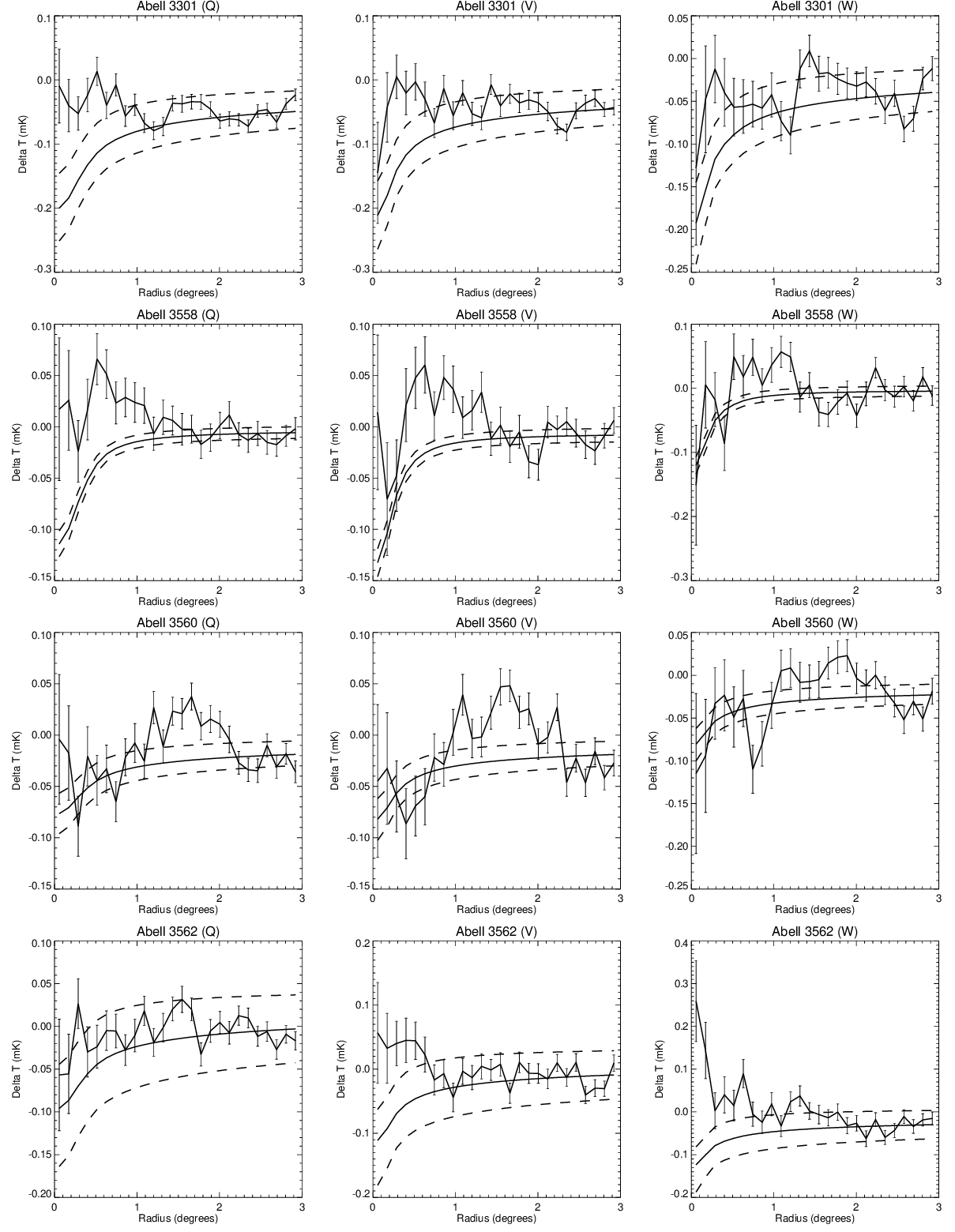}
\vspace{0.0mm}
\end{center}
\vspace{0.0cm}
\end{figure}

\begin{figure}[H]
\begin{center}
\includegraphics[angle=0,width=6in]{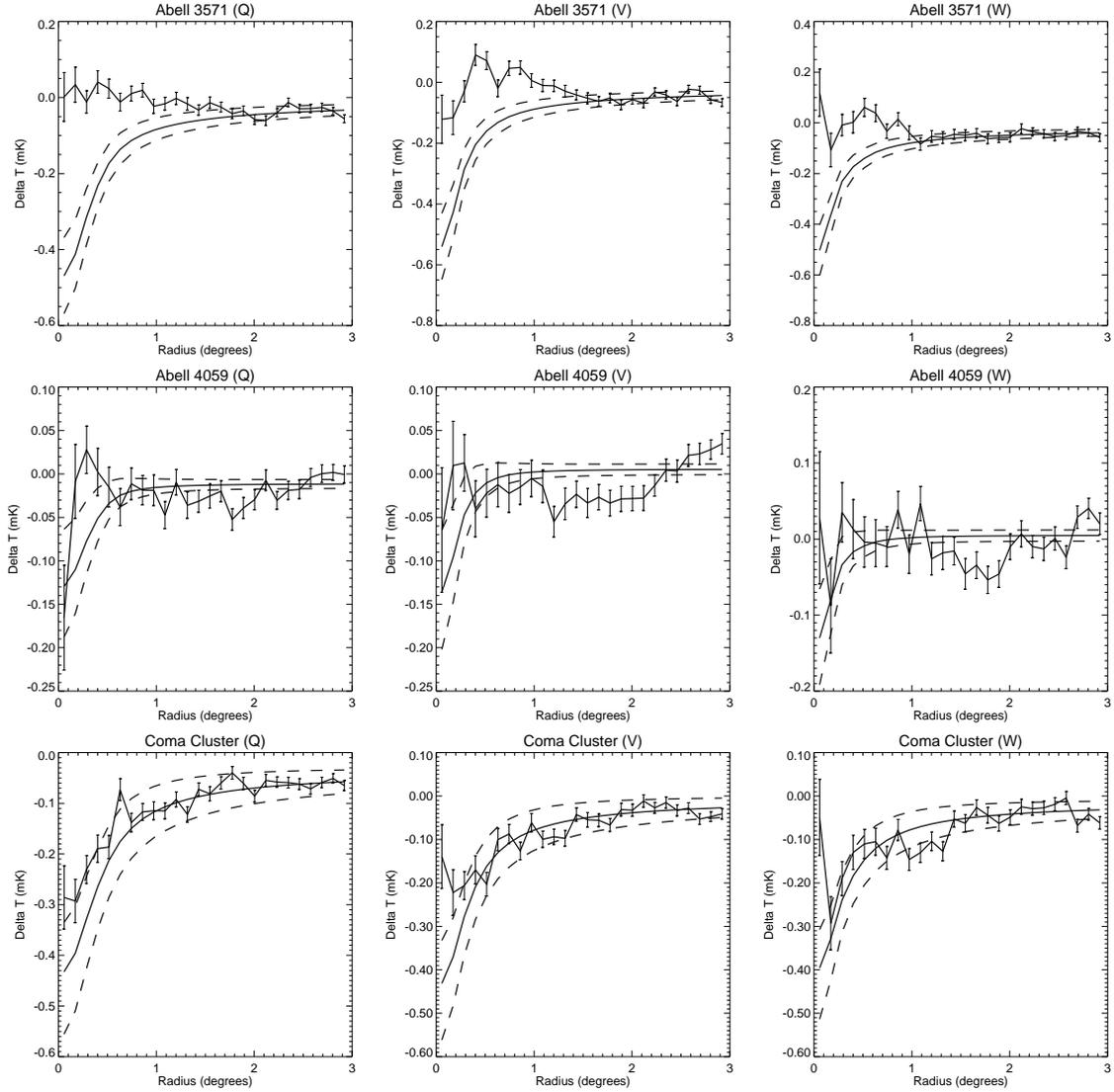}
\vspace{0.0mm}
\end{center}
\vspace{-3cm}
\caption{WMAP Q, V, and W band
radial profiles of the CMB temperature deviation (from
the all sky mean value) as averaged over concentric annuli
centered at the positions of the 31 clusters employed in
our analysis.  The data were plotted after subtraction
of the contribution from the CMB dipole and quadrupole anisotropy.
The solid line gives the SZE temperature decrement profile expected
from the hot ICM with properties determined by X-ray observations
(including the ROSAT $\beta$-model of Figure 1) and displayed
in Table 1.  Dashed lines mark the 90 \% error margin on the
prediction, based upon uncertainties in the observed hot ICM parameters.
The continuum level for the predicted SZE profile is fixed by
aligning it with the average temperature deviation in
the outermost (2$^\circ$ - 3$^\circ$) annuli.  This is a reliable procedure
because, according to the analysis of random fields the large annuli
averages stabilize to values near zero, see Figure 8.
\label{wmap}}
\vspace{0.0cm}
\end{figure}

\begin{figure}[H]
\begin{center}
\includegraphics[angle=0,width=4in]{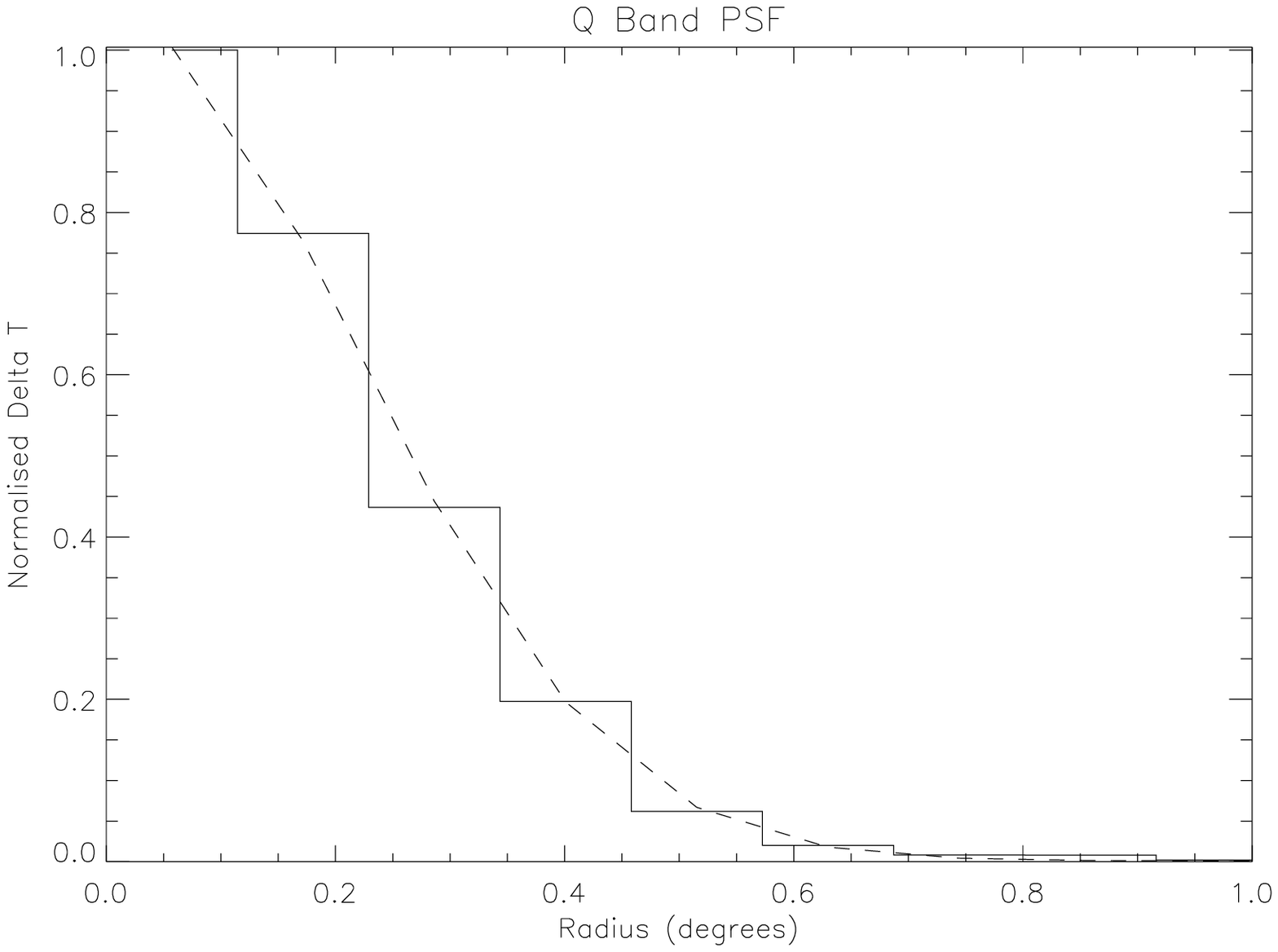}
\includegraphics[angle=0,width=4in]{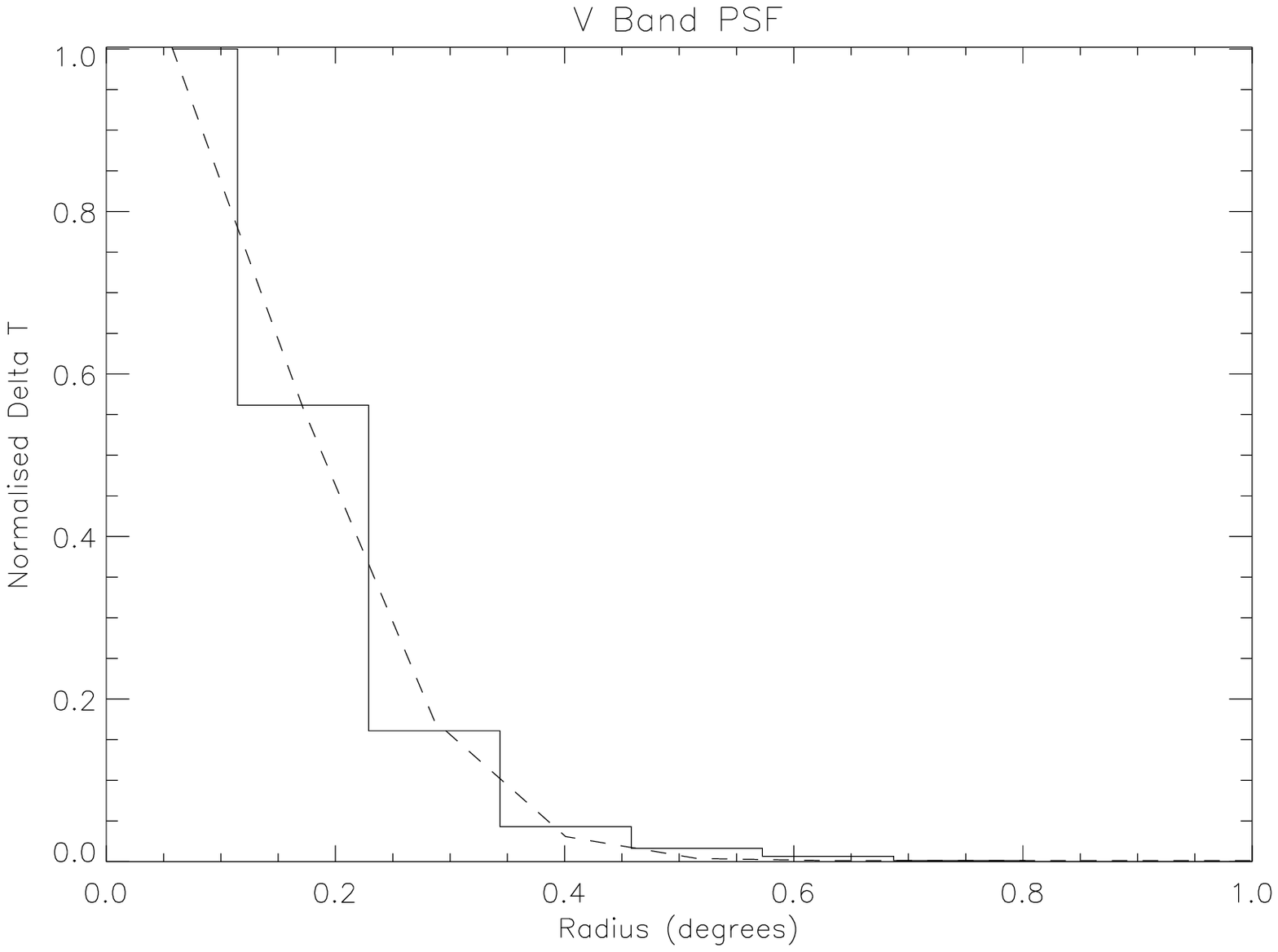}
\includegraphics[angle=0,width=4in]{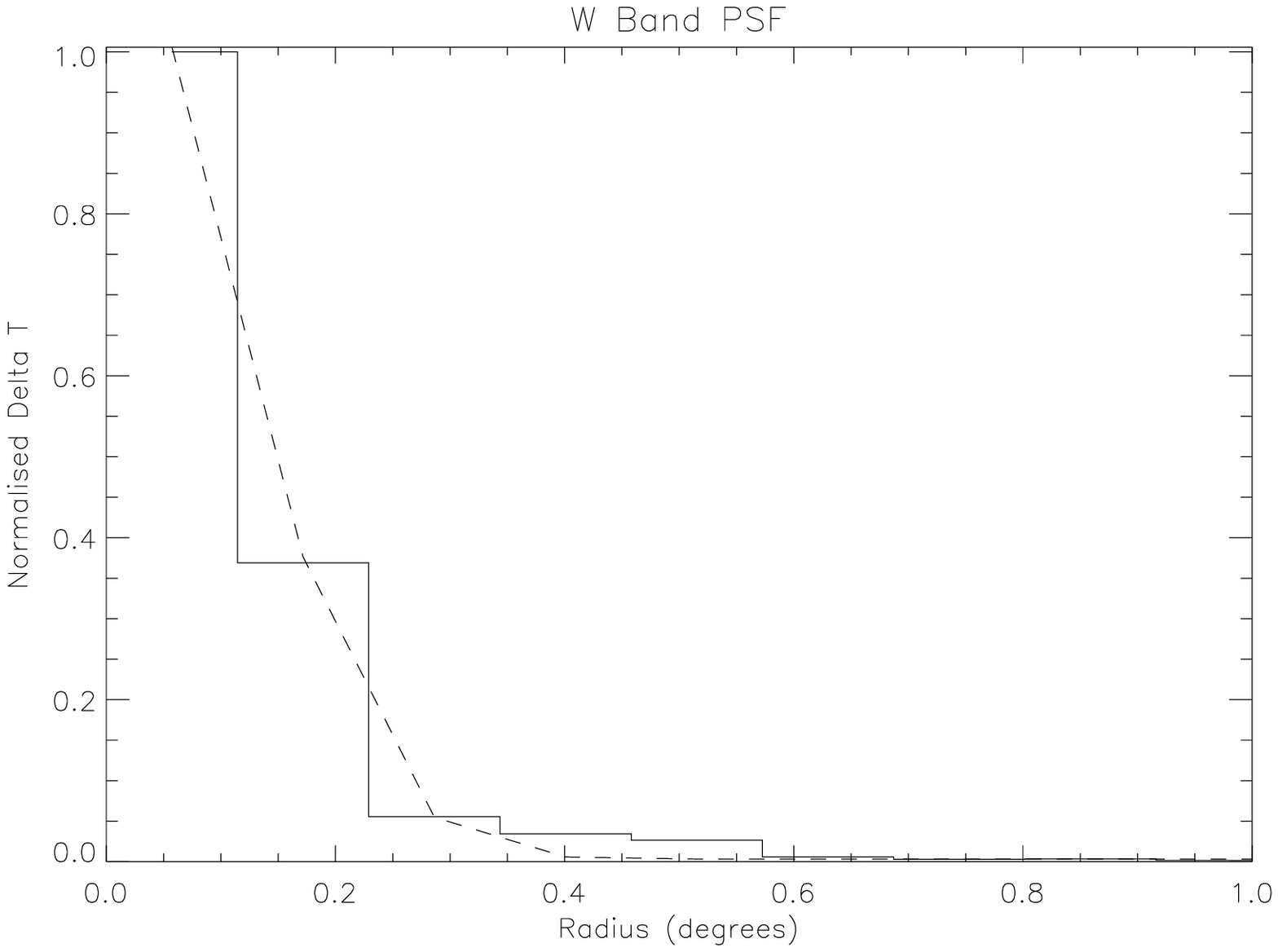}
\vspace{0.0mm}
\end{center}
\caption{Gaussian fits to the average profile of 15 point
sources, to determine the WMAP point spread function for the
Q, V, and W bands.  The best-fit model yielded $\sigma =$ 0.220$^\circ$ for
Q, 0.150$^\circ$ for V, and 0.115$^\circ$ for W.  The values are in good
agreement with those of Myers et al (2004).
\label{wmap-psf}}
\vspace{0.0cm}
\end{figure}

\begin{figure}[H]
\begin{center}
\includegraphics[angle=0,width=4in]{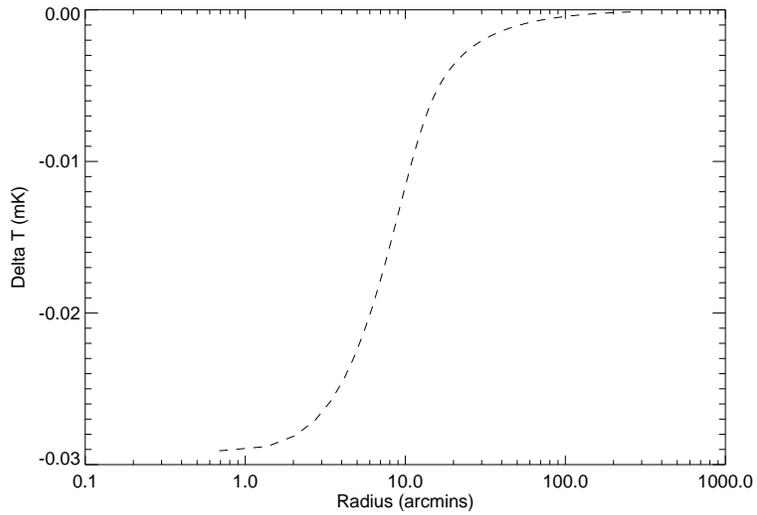}
\vspace{0.0mm}
\end{center}
\caption{Our PSF for the WMAP W band (as derived from Figure 3)
convolved with the hot ICM model $\beta =$ 0.74, $\theta_c =$ 1.5 arcmin,
and amplitude $\Delta T =$ -0.083 K.  The model parameters are chosen
for direct comparison with Figure 1(c) of Myers et al (2004), which 
displays the temperature profile after the same model was convolved with
their estimate of the W-band PSF.  It can be seen that the two curves
resemble each other very closely.  This checks the correctness of
our convolution routine.
\label{shanks}}
\vspace{0.0cm}
\end{figure}

\begin{figure}[H]
\begin{center}
\includegraphics[angle=0,width=4in]{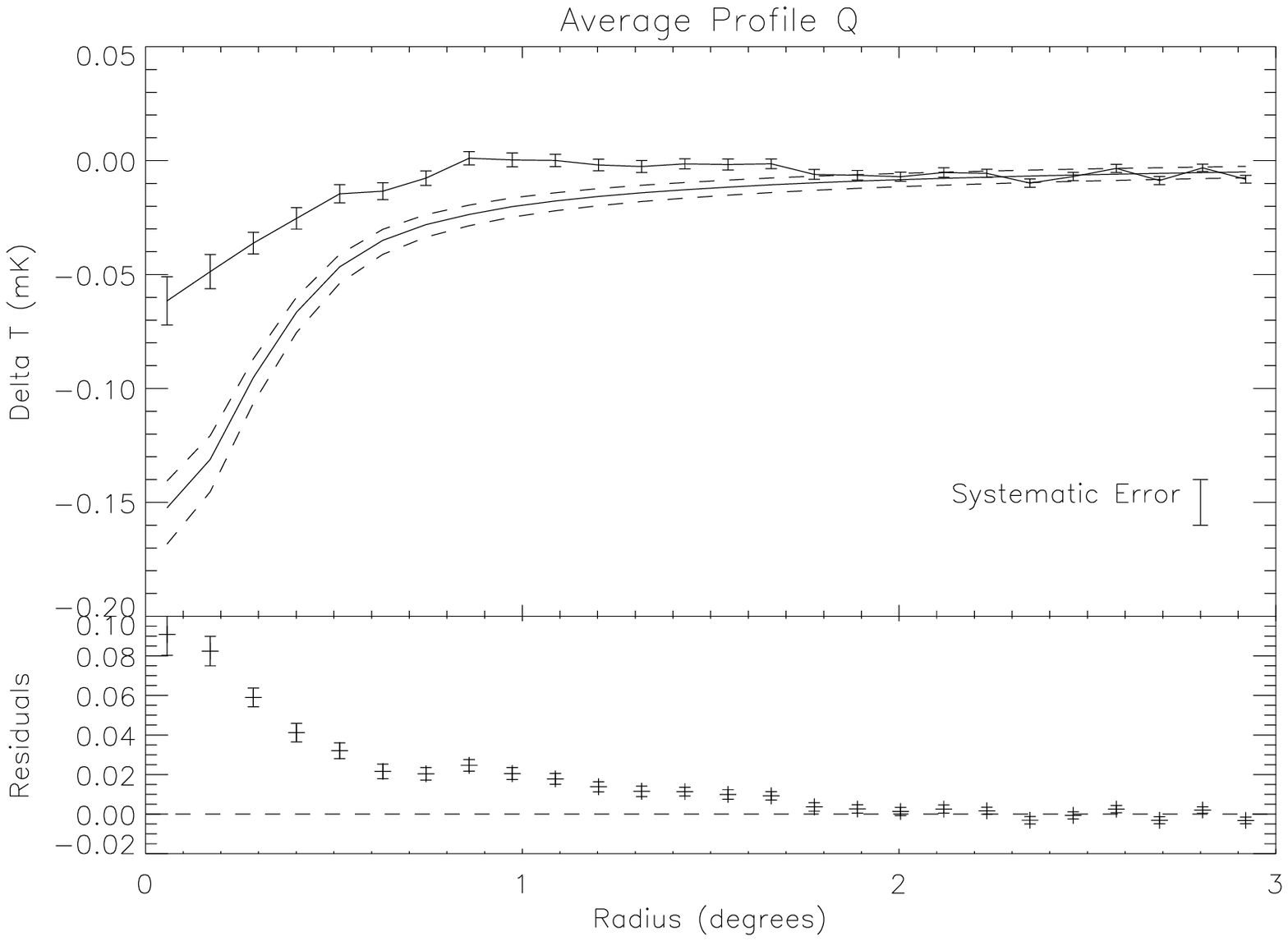}
\includegraphics[angle=0,width=4in]{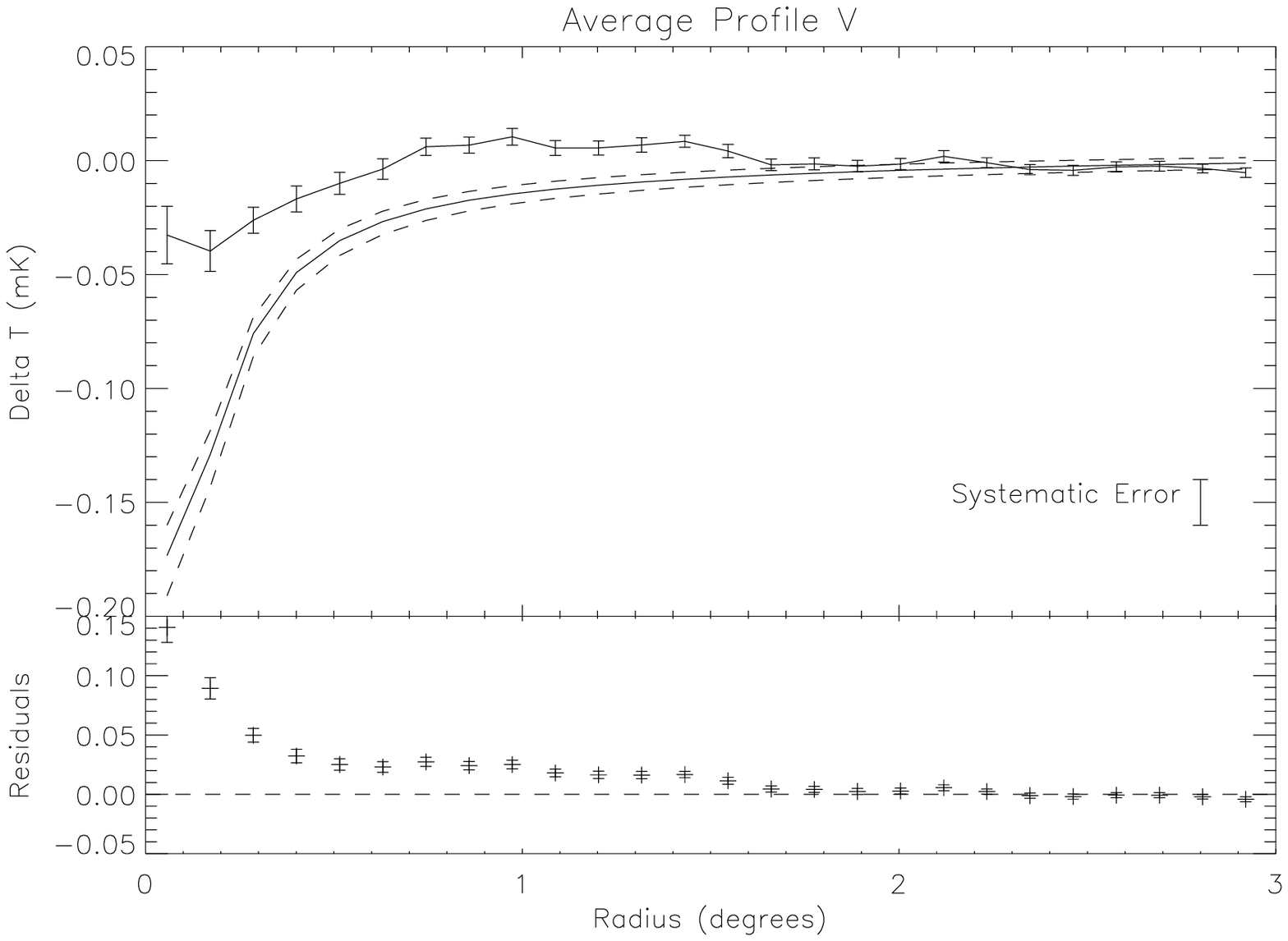}
\includegraphics[angle=0,width=4in]{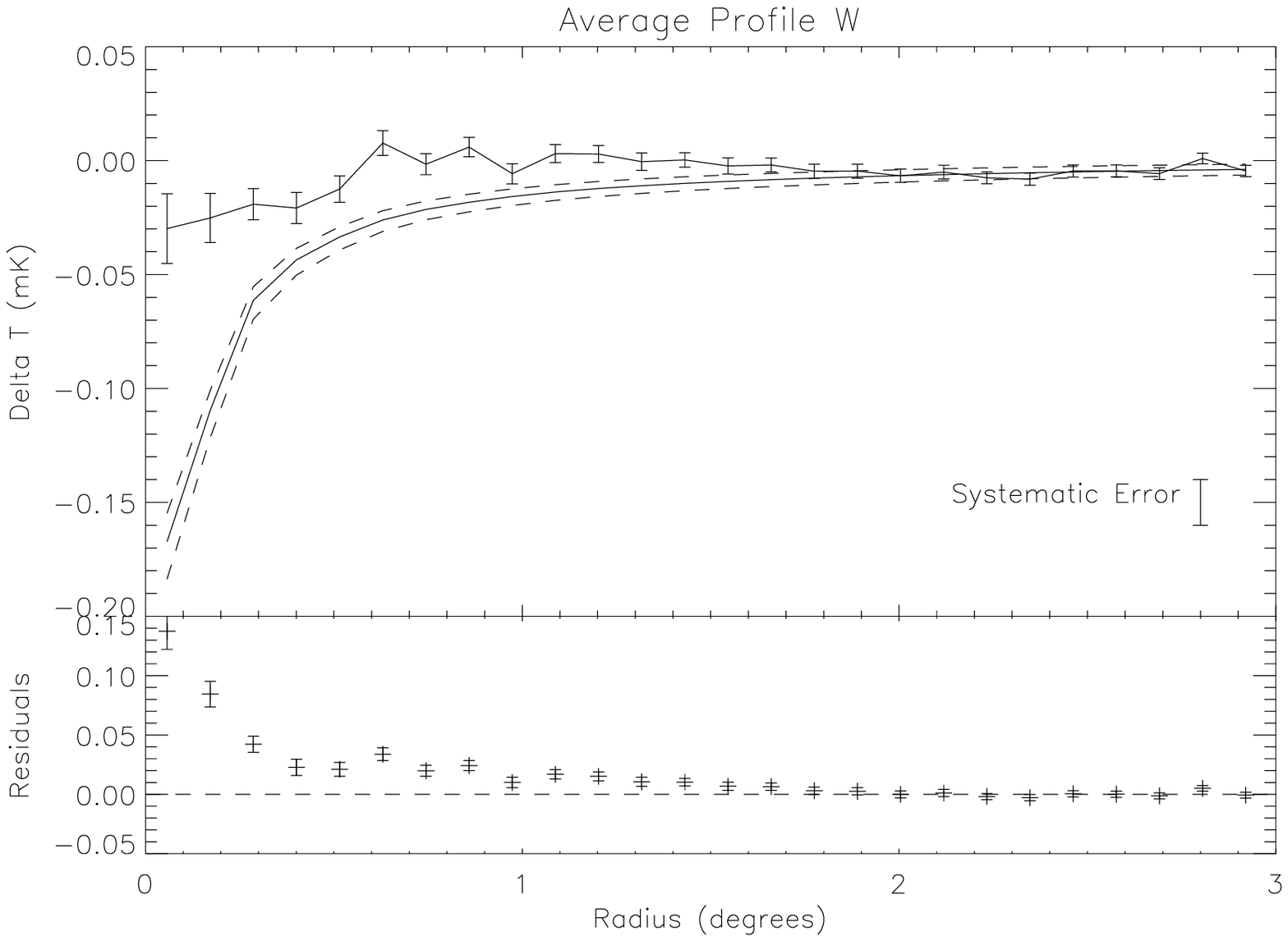}
\vspace{-1cm}
\end{center}
\caption{The average WMAP observed and predicted radial profile
for the 31
clusters of our sample.  An individual error bar for each bin
depicts the random uncertainty (i.e. WMAP antenna noise), while the
systematic 1-$\sigma$ error plotted on the side of the graph depicts
the residual  large scale correlated variation  in the central
1$^\circ$ region of the 33
co-added random fields of Figure 8.
The continuum of the prediction curve is fixed
by alignment with the 2$^\circ$ -- 3$^\circ$ data.
This alignment procedure is justified by
Figure 8, where it was shown that when 33 random fields were co-added
the 2$^\circ$ -- 3$^\circ$ annulus
exhibits high stability.
\label{sze-prediction}}
\vspace{0.0cm}
\end{figure}

\begin{figure}[H]
\begin{center}
\includegraphics[angle=0,width=4in]{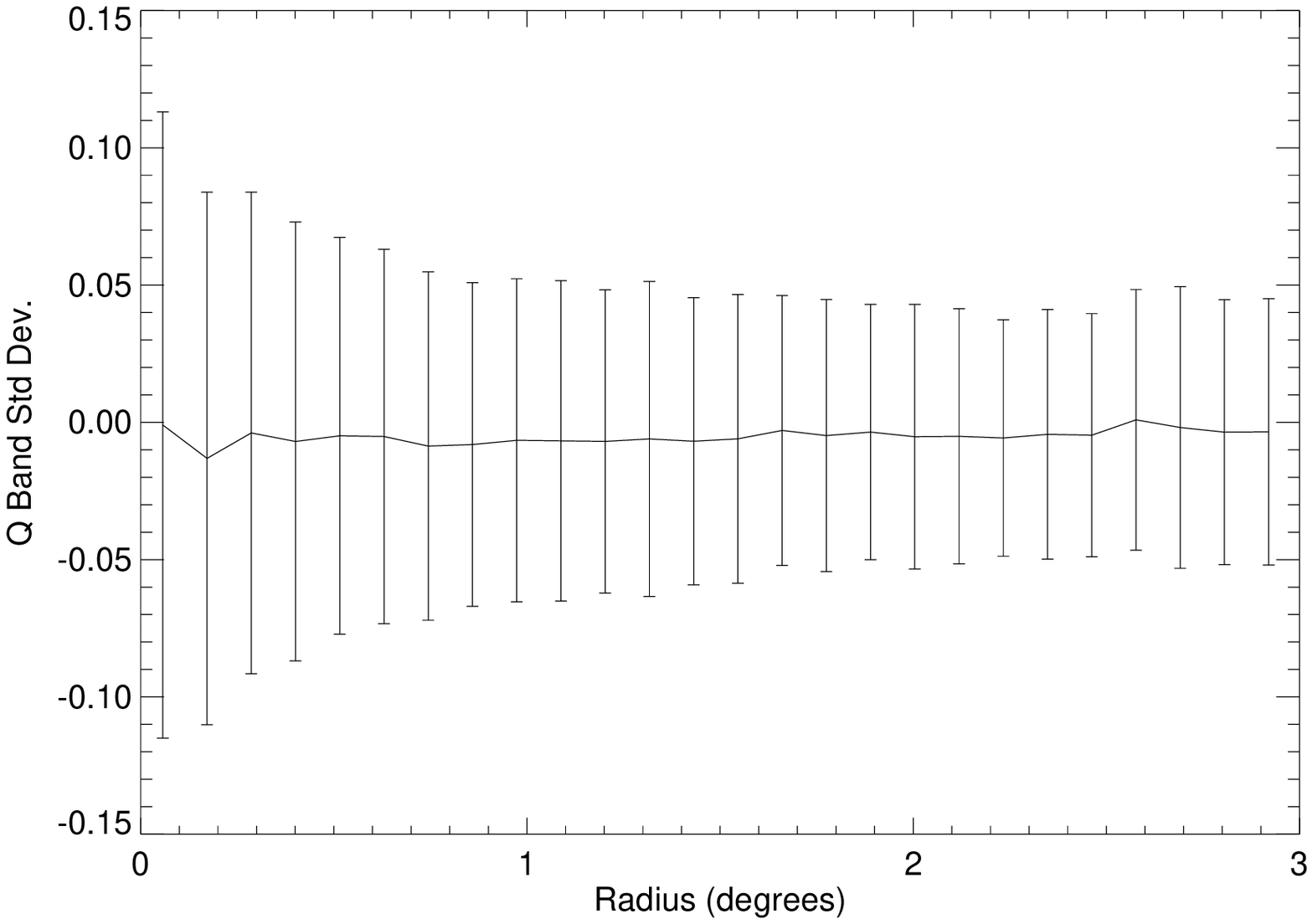}
\includegraphics[angle=0,width=4in]{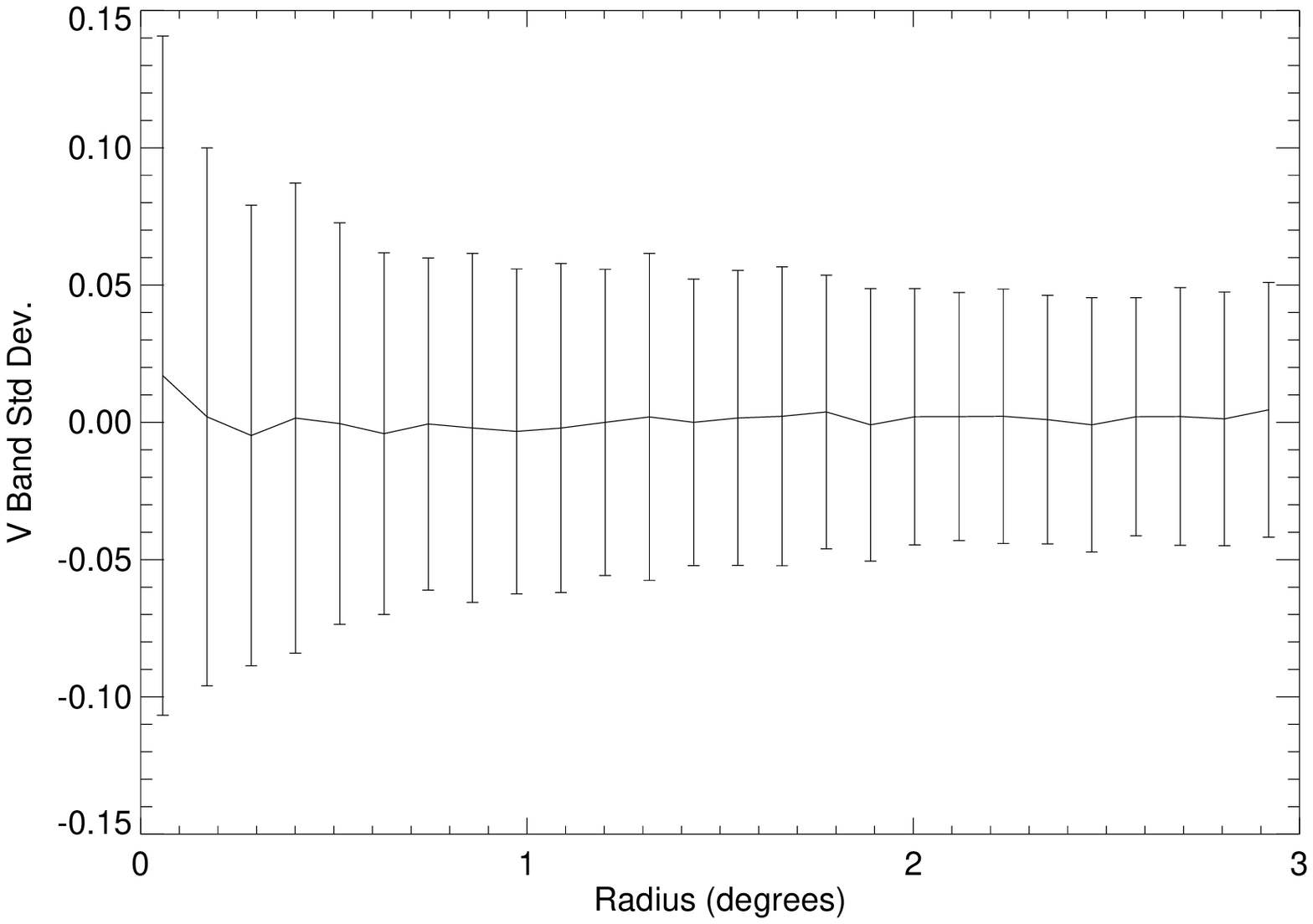}
\includegraphics[angle=0,width=4in]{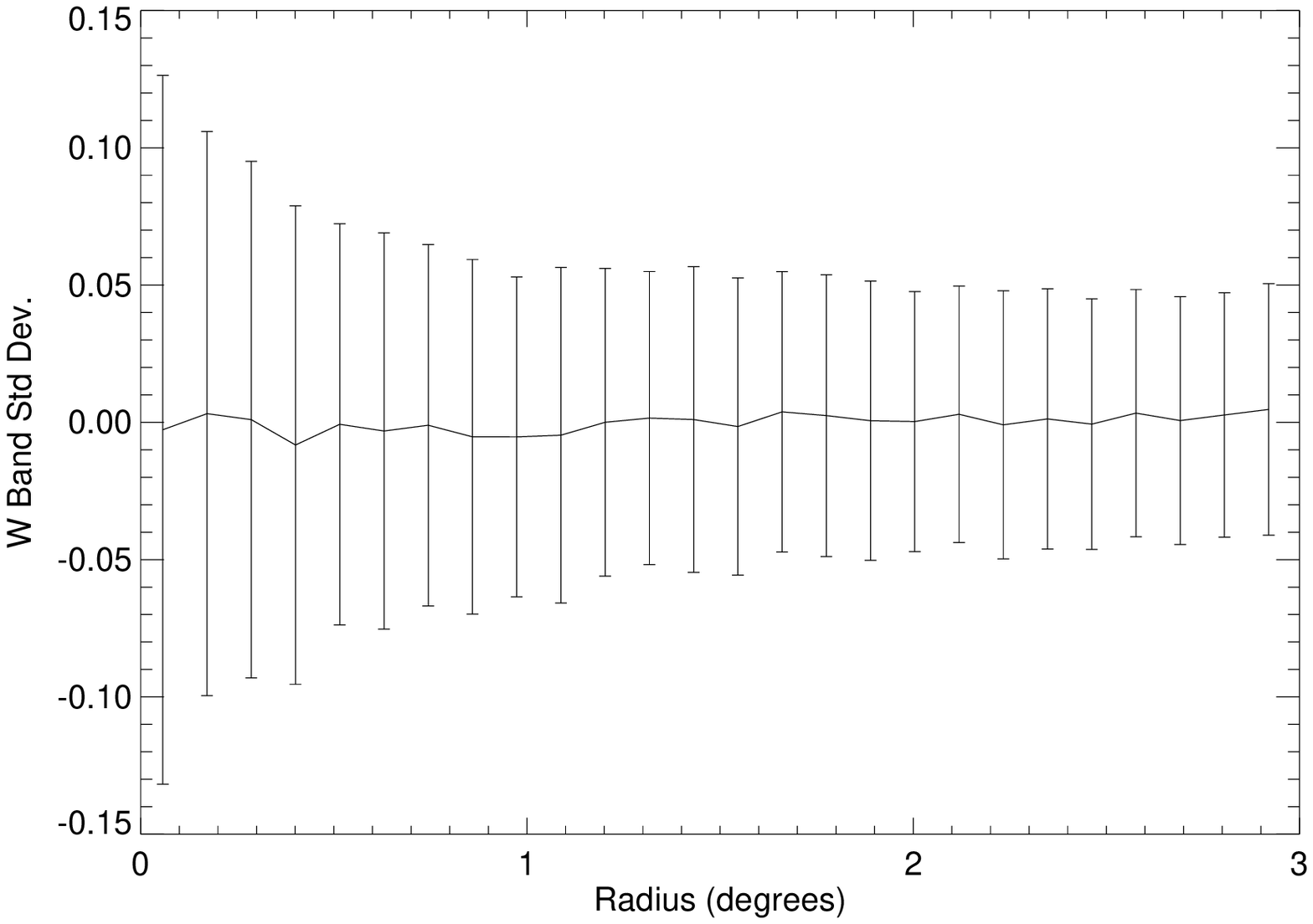}
\vspace{0.0mm}
\end{center}
\caption{The r.m.s. field-to-field variation of CMB temperature
at a given radial interval, and for the Q, V, W bands of WMAP, as obtained
by comparing the radial profile 100 random pointings.
\label{random-stdev}}
\vspace{0.0cm}
\end{figure}

\begin{figure}[H]
\begin{center}
\includegraphics[angle=0,width=6in]{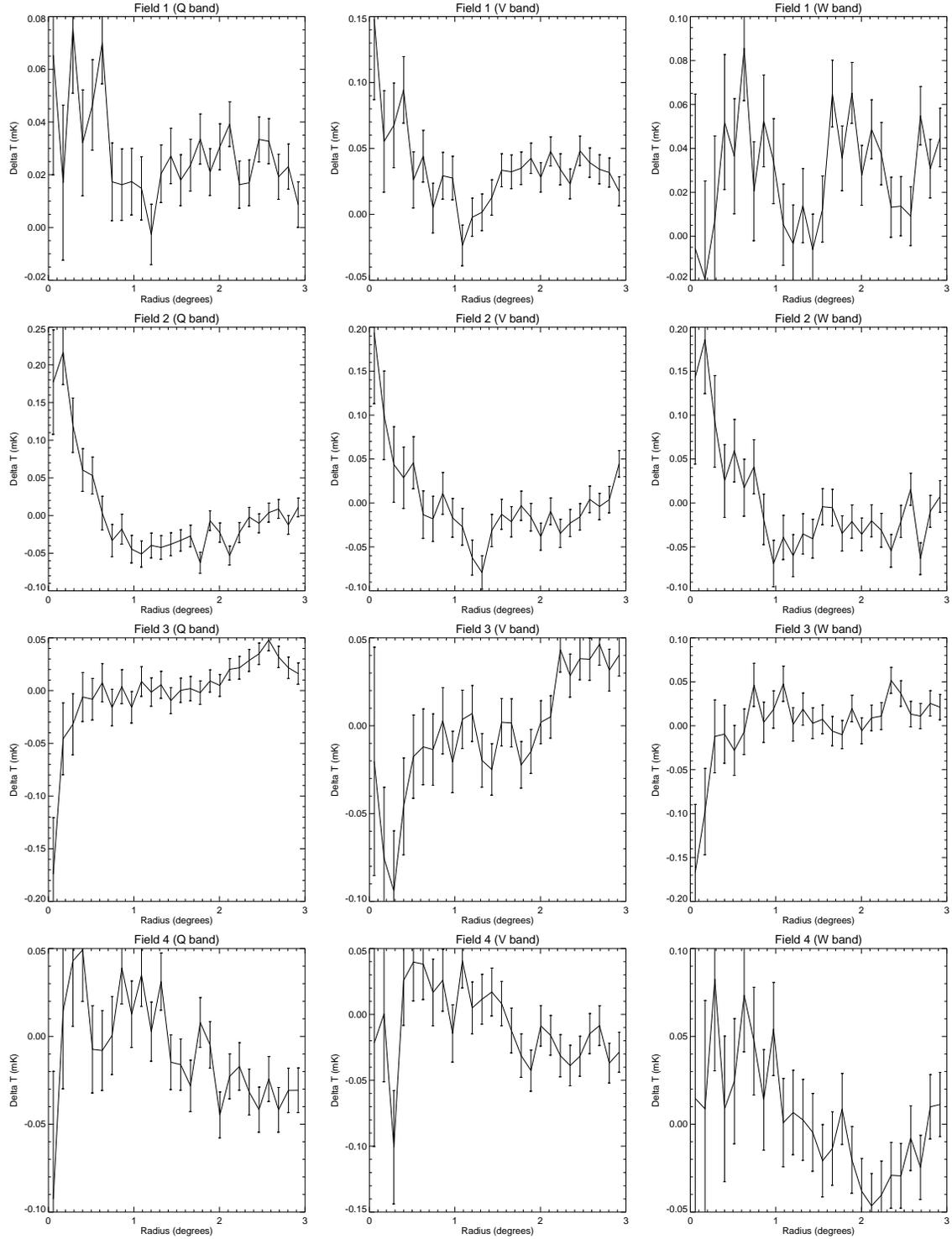}
\vspace{0.0mm}
\end{center}
\caption{The WMAP Q, V, and W band radial profile (of CMB
temperature deviation) centered at four randomly
chosen positions on the sky.
\label{random-single}}
\vspace{0.0cm}
\end{figure}

\begin{figure}[H]
\begin{center}
\includegraphics[angle=0,width=4in]{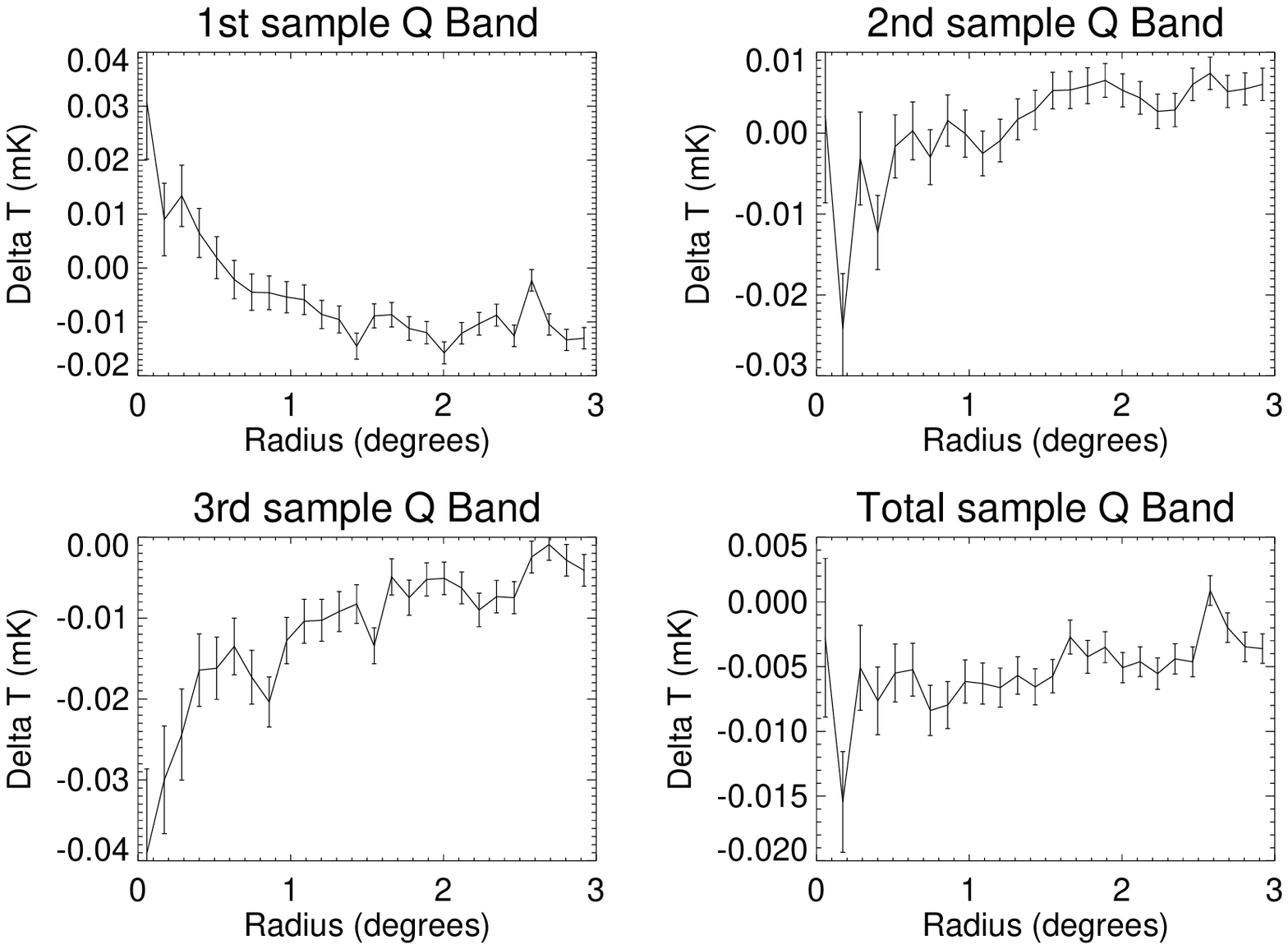}
\includegraphics[angle=0,width=4in]{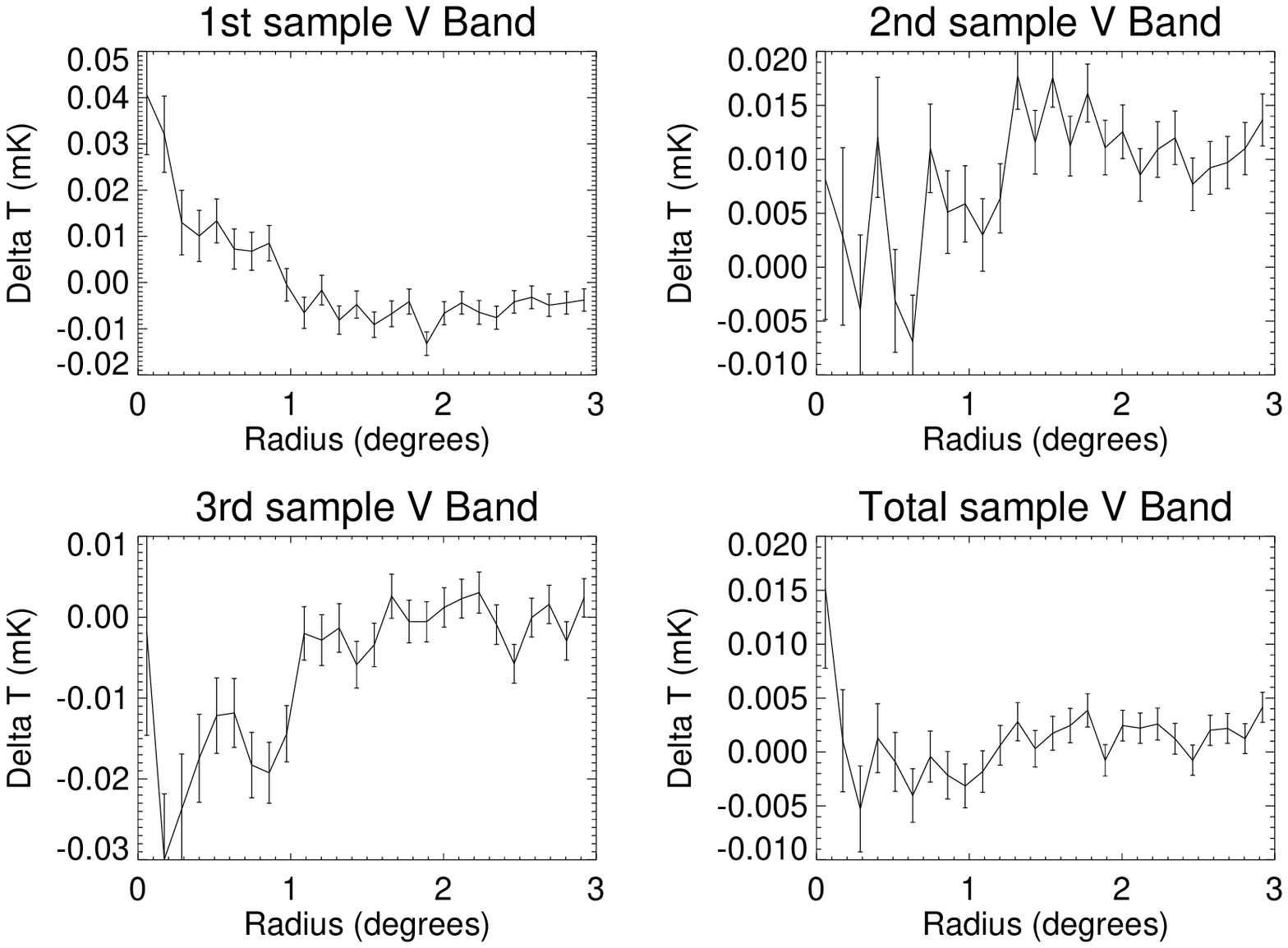}
\includegraphics[angle=0,width=4in]{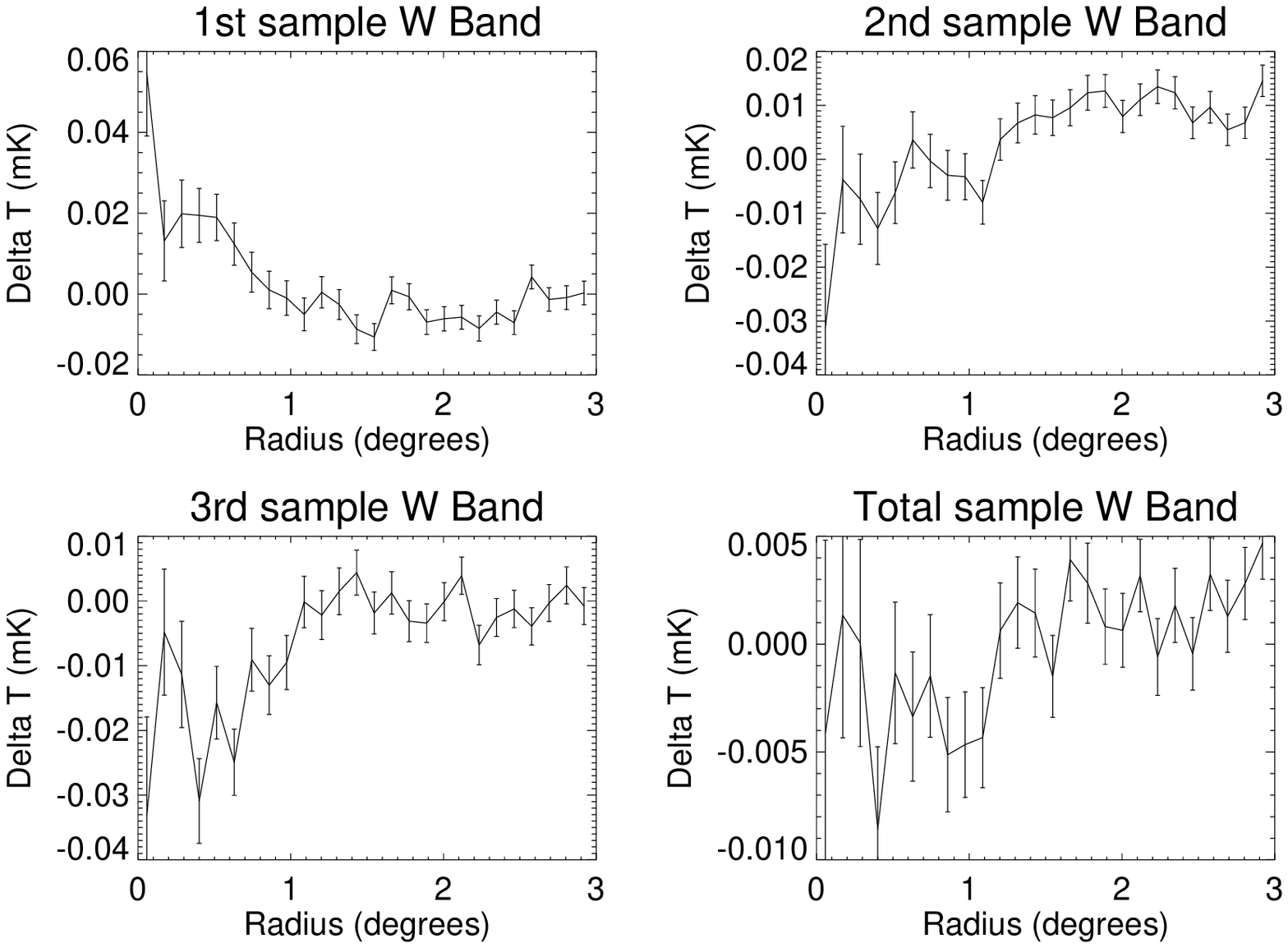}
\vspace{0.0mm}
\end{center}
\caption{The WMAP Q, V, and W band radial profile (of CMB
temperature deviation).  For each band the first three graphs show
the average profile of  33 fields, each centered at a randomly
chosen position on the sky, while the last (bottom right) graph
shows the average profile of 100 such fields - here the r.m.s.
scatter of the data confirms the correctness of the size of the
error bars.
\label{random}}
\vspace{0.0cm}
\end{figure}

\begin{figure}[H]
\begin{center}
\includegraphics[angle=0,width=4in]{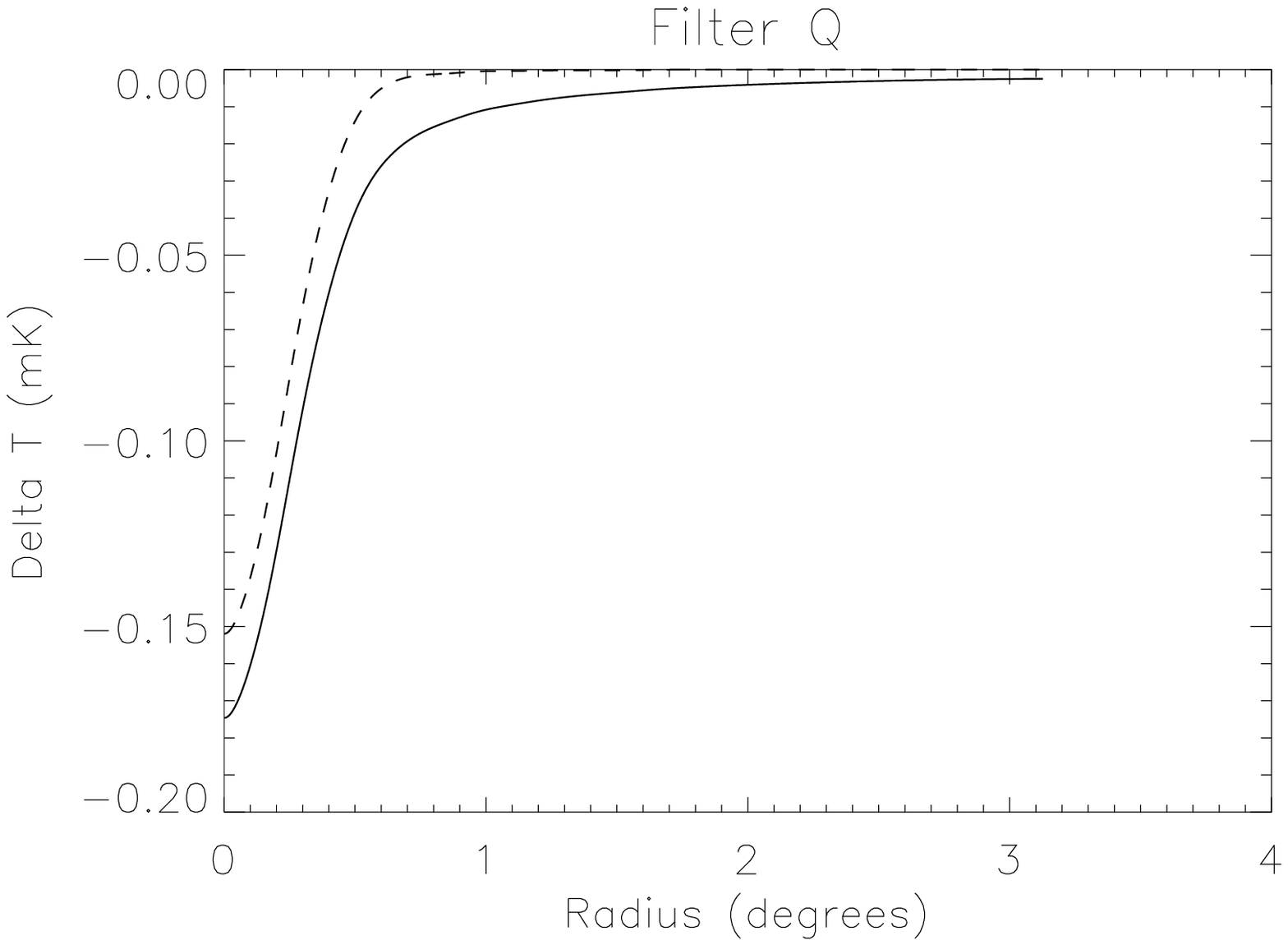}
\includegraphics[angle=0,width=4in]{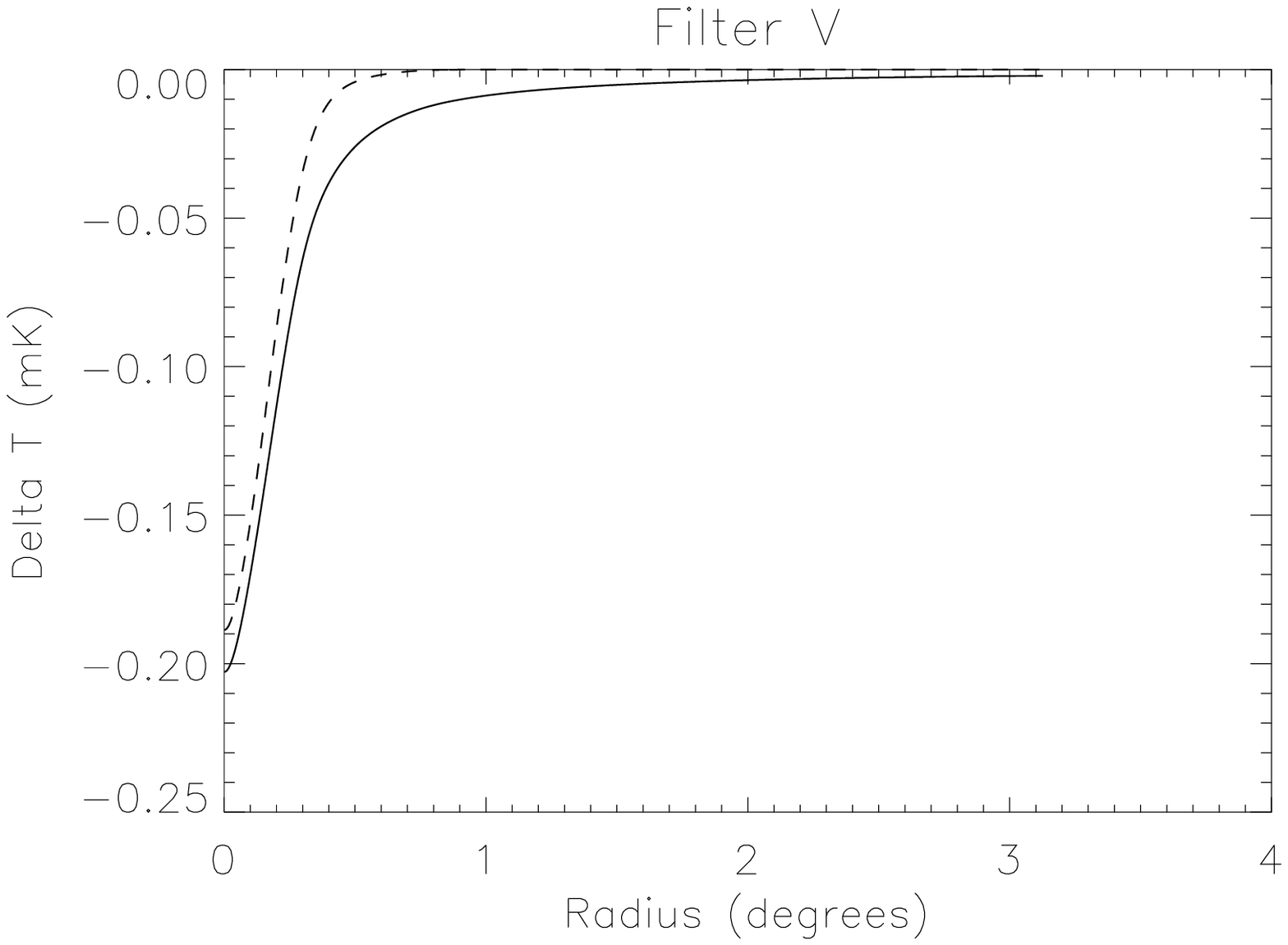}
\includegraphics[angle=0,width=4in]{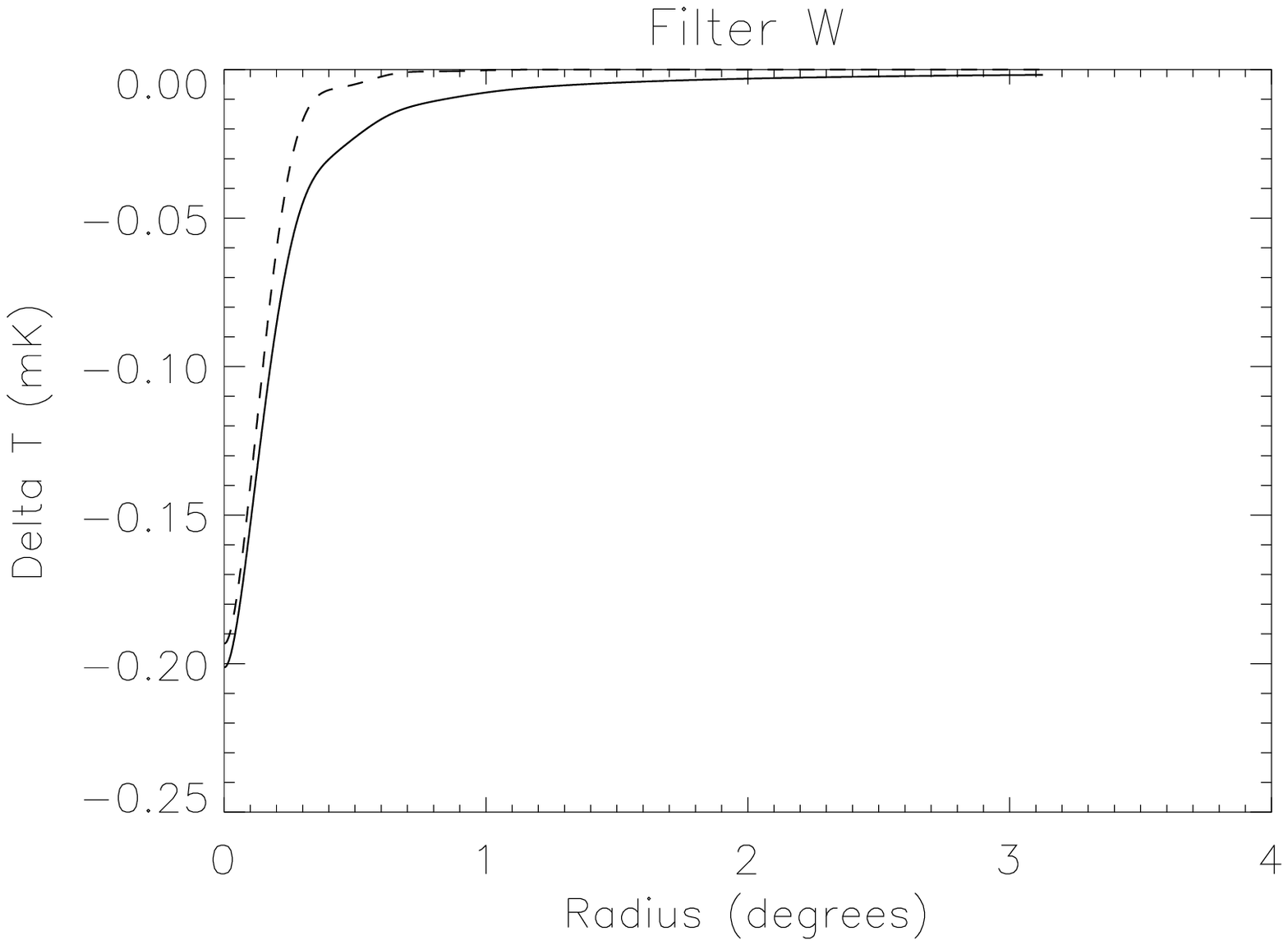}
\vspace{0.0mm}
\end{center}
\caption{A comparison between the SZE predicted profile
of two versions of the 
$\beta$ model, after
each profile is convolved with the WMAP PSF (for the appropriate filter).
Both versions use exactly the same model parameters as starting point, except 
one invokes the full model of Eq. (4) with $\theta_c =$ 2 arcmin and
$\beta =$ 2/3 (solid line), while the other (dashed
line) invokes a truncation of the
model at $\theta =$ 10 arcmin, i.e. Eq. (6).
\label{convergence}}
\vspace{0.0cm}
\end{figure}

\begin{figure}[H]
\begin{center}
\includegraphics[angle=0,width=4in]{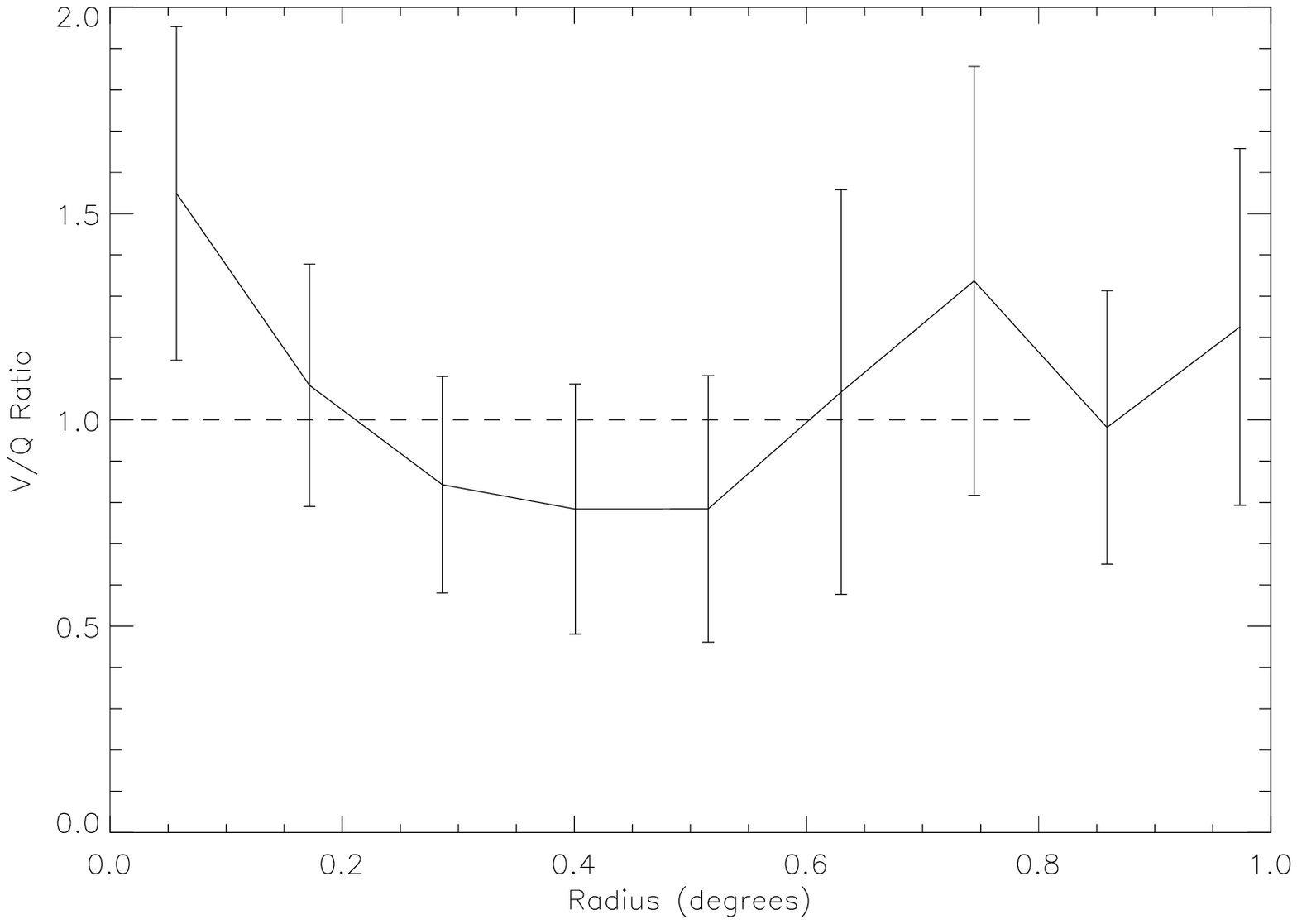}
\includegraphics[angle=0,width=4in]{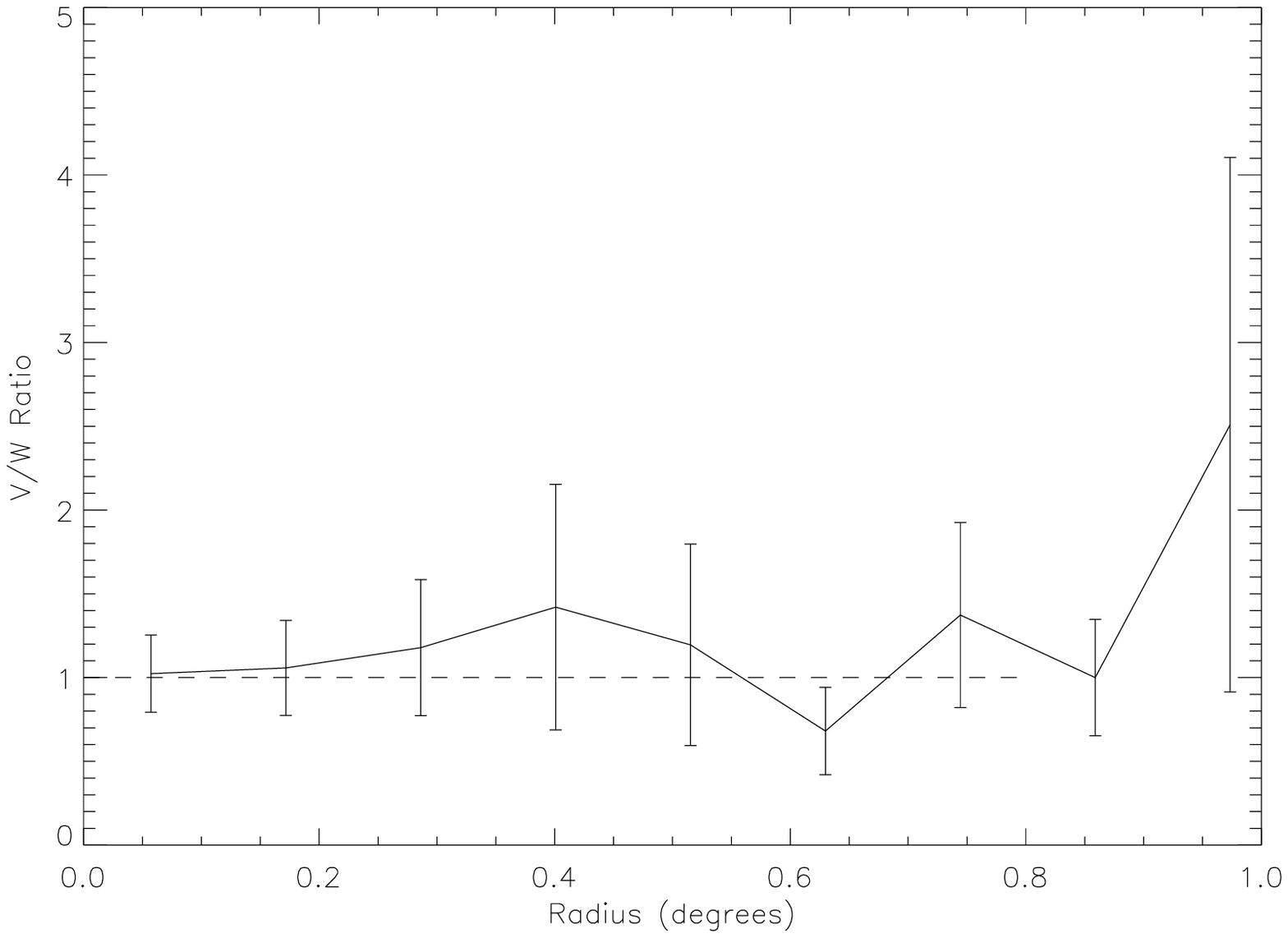}
\includegraphics[angle=0,width=4in]{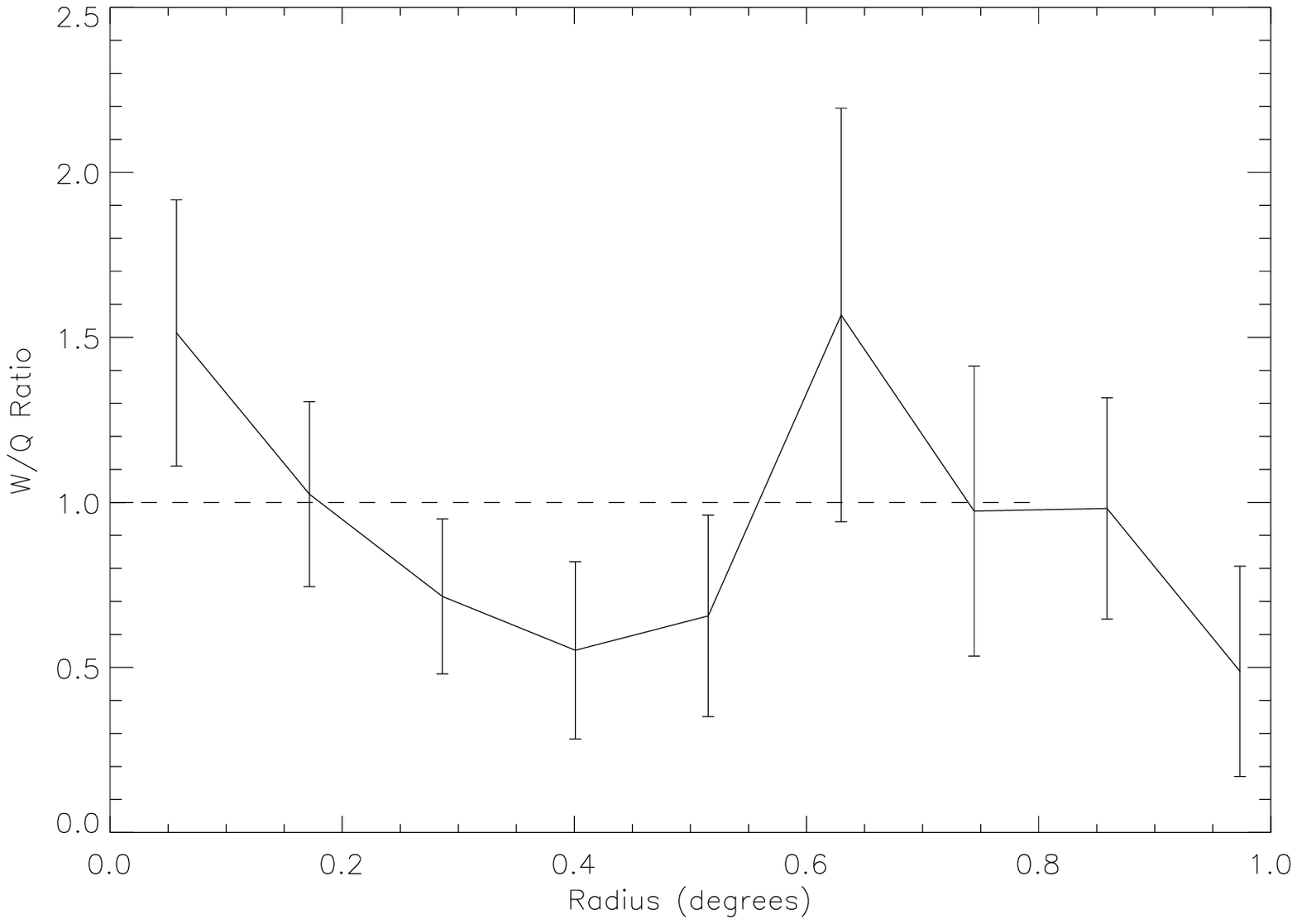}
\vspace{0.0mm}
\end{center}
\caption{Radial profile of the filter
ratio of the temperature discrepancy between SZE model and
data, for the 31 co-added cluster fields of Figure 5.  First plot
gives V:Q ratio, second is V:W, and third is W:Q.  If the
missing flux responsible for the discrepany has a black body
spectrum, the filter ratios will equal unity.
\label{ratio}}
\vspace{0.0cm}
\end{figure}

\begin{figure}[H]
\begin{center}
\includegraphics[angle=0,height=7in,width=6in]{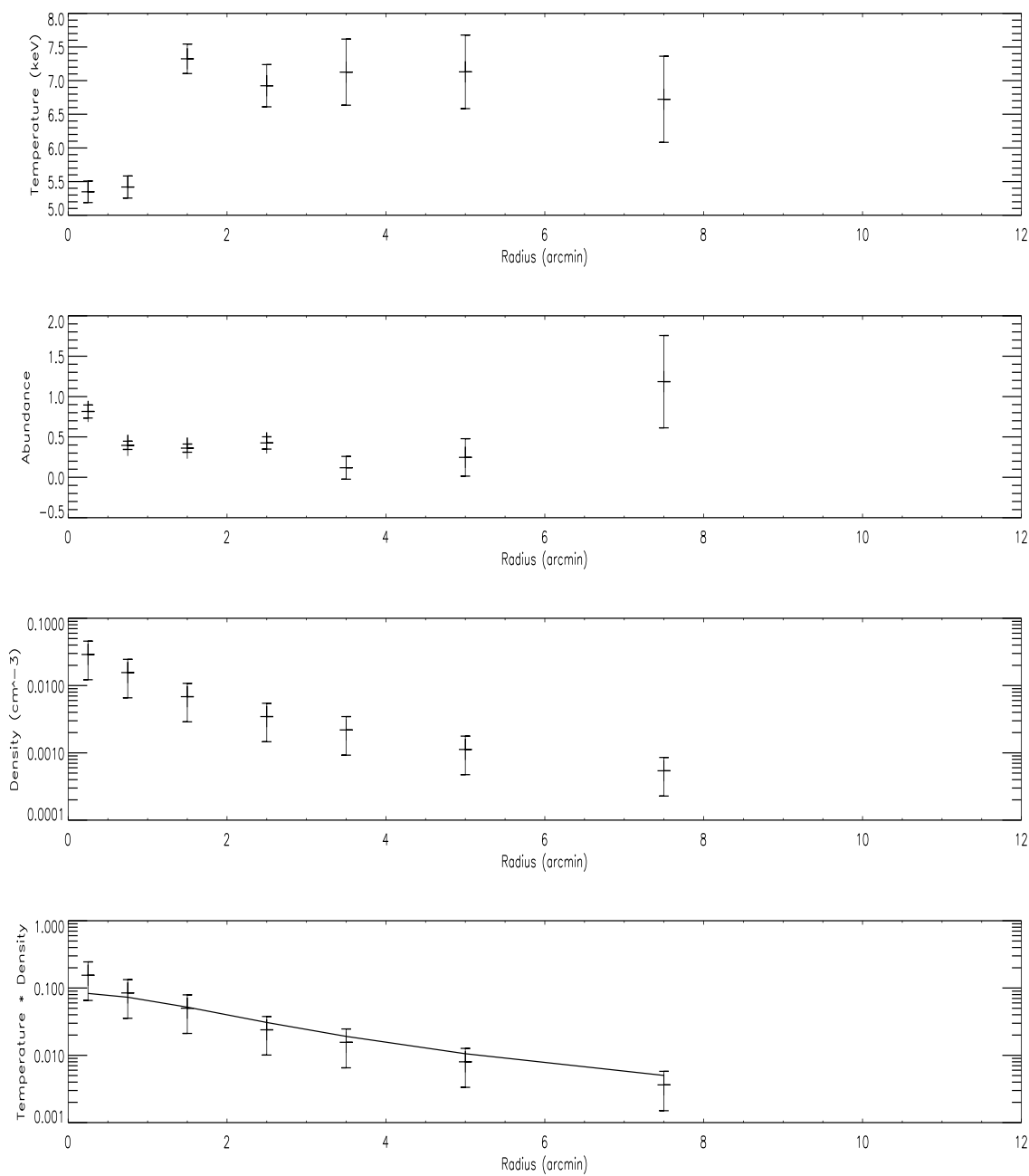}
\vspace{0.0mm}
\end{center}
\caption{XMM-Newton deprojected best hot ICM parameters of
A2029.  The plotted values are averaged over all three instruments
MOS1, MOS2, and PN.  In the bottom graph the solid line represents
the $\beta$-model as fitted to the ROSAT data outside
the `cooling flow' region, with the density profile then  extrapolated
inwards whilst assuming
isothermality for all radii at the temperature given by Table 1
(i.e. ignoring the cooling core, which is exactly
our procedure when we derived the predicted 
SZE flux from the parameters of Table 1).   
The reason for plotting the product of deprojected density and temperature
is because the X-ray SZE prediction depends directly on this quantity.
It can be seen that our method yields good agreement with XMM-Newton data,
even though this cluster exhibits the worst discrepancy between X-ray
and WMAP measurements.
\label{xmm}}
\vspace{0.0cm}
\end{figure}

\begin{figure}[H]
\begin{center}
\includegraphics[angle=0,height=7in,width=6in]{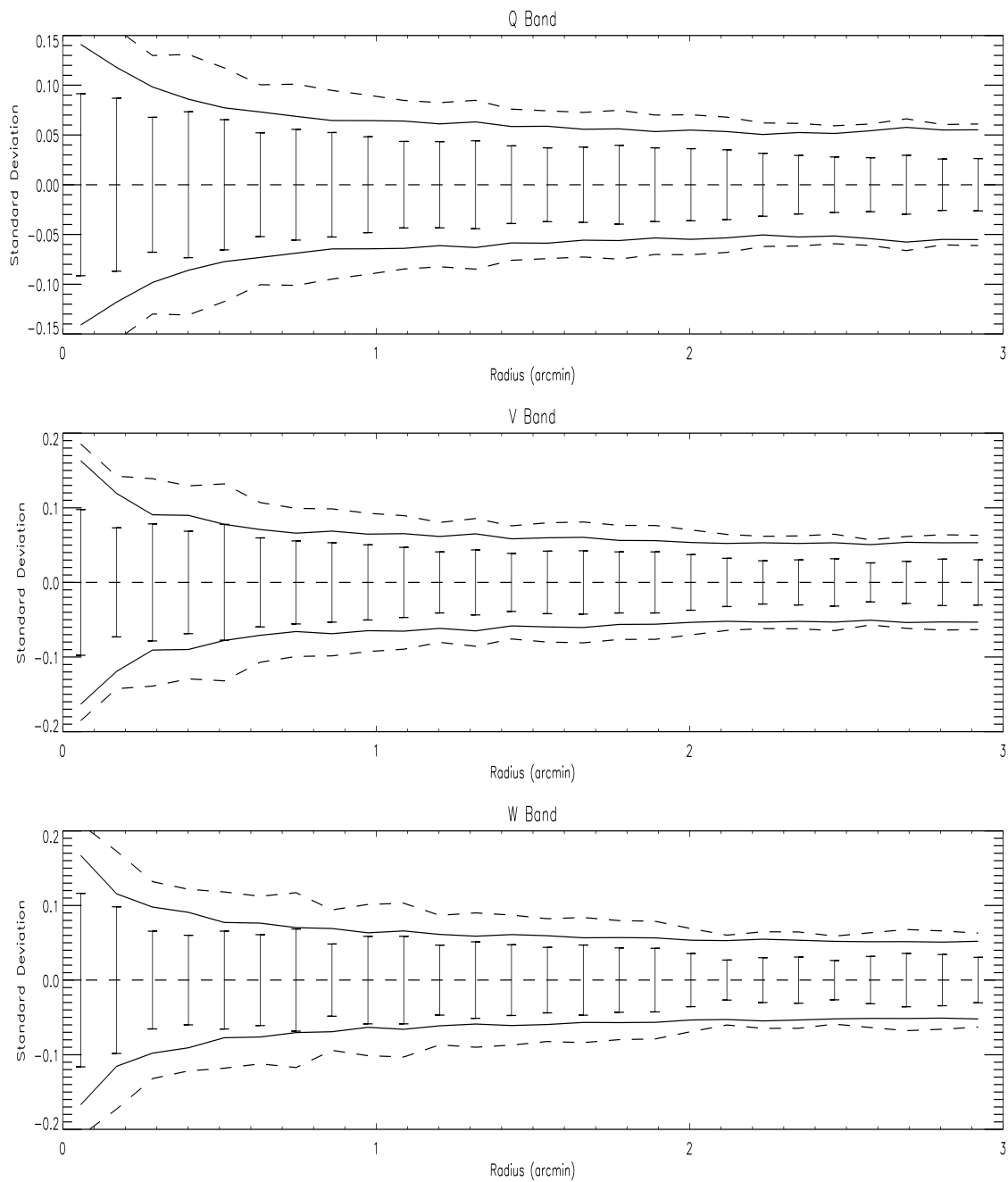}
\vspace{0.0mm}
\end{center}
\caption{Cluster field-to-field CMB temperature variation as seen in
the three WMAP passbands at each radial bin.  The solid line represents
the level expected from the combined effect of
natural blank field (i.e. primary CMB) variation and the SZE
contrasts among clusters.  The dashed line represents the level expected
if radio sources with average brightness comparable to the SZE decrement
are present in each cluster with no correlation between such sources and
the properties of the hot ICM within the same cluster.  The fact that
both solid and dashed lines are above the level of the data fluctuations
is evidence for the presence of {\it neither} the SZE, nor radio
contaminations at the level that can offset the SZE.
\label{rms}}
\vspace{0.0cm}
\end{figure}

\end{document}